\documentclass[useAMS,usenatbib]{mn2e}
\usepackage{epsfig,amsmath,natbib}
\usepackage{color,subfigure}
\usepackage{./style/natbibmnfix}
  
\usepackage{mathrsfs,amssymb,amstext}
\usepackage{ulem}
\usepackage{url}
\usepackage{bm}
\usepackage{adjustbox}

\def\be{\begin{equation}} 
\def\ee{\end{equation}} 
\def\ba{\begin{eqnarray}} 
\def\ea{\end{eqnarray}}

\def\kms{\,{\rm {km\, s^{-1}}}}

\def\HI{\hbox{H~$\scriptstyle\rm I\ $}}

\def\gsim{\lower.5ex\hbox{\gtsima}} 
\def\lsim{\lower.5ex\hbox{\ltsima}} \def\gtsima{$\; \buildrel > \over 
\sim \;$} \def\ltsima{$\; \buildrel < \over \sim \;$} \def\prosima{$\; 
\buildrel \propto \over \sim \;$} \def\gsim{\lower.5ex\hbox{\gtsima}} 
\def\lsim{\lower.5ex\hbox{\ltsima}} 
\def\simgt{\lower.5ex\hbox{\gtsima}} 
\def\simlt{\lower.5ex\hbox{\ltsima}} 
\def\simpr{\lower.5ex\hbox{\prosima}} \def\la{\lsim} \def\ga{\gsim} 
  
\def\gtsima{$\; \buildrel > \over \sim \;$} 
\def\ltsima{$\; \buildrel < \over \sim \;$} 
\def\gsim{\lower.5ex\hbox{\gtsima}} 
\def\lsim{\lower.5ex\hbox{\ltsima}} 
\def\simgt{\lower.5ex\hbox{\gtsima}} 
\def\simlt{\lower.5ex\hbox{\ltsima}} 
\def\simpr{\lower.5ex\hbox{\prosima}} 
\def\la{\lsim} 
\def\ga{\gsim}

\def\E3{{\cal E}_{\rm g}^{III}}

\def\x12{x_{1/2}} 

\def\gtsima{$\; \buildrel > \over \sim \;$}
\def\ltsima{$\; \buildrel < \over \sim \;$}
\def\gsim{\lower.5ex\hbox{\gtsima}}
\def\lsim{\lower.5ex\hbox{\ltsima}}
\def\simgt{\lower.5ex\hbox{\gtsima}}
\def\simlt{\lower.5ex\hbox{\ltsima}}
\def\simpr{\lower.5ex\hbox{\prosima}}

\def\CII{\hbox{[C$\scriptstyle\rm II$]\,}}

\def\OI{\hbox{[O$\scriptstyle\rm I$]\,}}
\def\OIII{\hbox{[O$\scriptstyle\rm III$]\,}}

\def\mean#1{\left< #1 \right>}

\title[\CII intensity mapping of high-$z$ galaxies]{Studying high-$z$ galaxies with \CII~intensity mapping} 
\author[Yue \& Ferrara]{
B. Yue$^{1}$ \& A. Ferrara$^{2,3}\thanks{E-mail: andrea.ferrara@sns.it}  $
\\
$^{1}$National Astronomical Observatories, Chinese Academy of Sciences, 20A, Datun Road, Chaoyang District, Beijing, 100101, China\\
$^{2}$Scuola Normale Superiore, Piazza dei Cavalieri 7, 56126 Pisa, Italy\\
$^{3}$Kavli Institute for the Physics and Mathematics of the Universe (WPI), University of Tokyo, Kashiwa 277-8583, Japan\\
}

\begin{document} 
 
\date{\today} 
 
\pagerange{\pageref{firstpage}--\pageref{lastpage}} \pubyear{2012} 
 
\maketitle 
 
\label{firstpage} 
\begin{abstract} 
We investigate the \CII line intensity mapping (IM) signal from galaxies in the Epoch of Reionization (EoR) to assess its detectability, the possibility to constrain the $L_{\rm CII}-{\rm SFR}$ relation, and to recover the \CII luminosity function (LF) from future experiments. By empirically assuming that ${\rm log}L_{\rm CII}={\rm log}A+\gamma {\rm SFR}\pm\sigma_L$,
we derive the \CII LF from the observed UV LF, and the \CII IM power spectrum. We study the shot-noise and the full power spectrum separately. Although, in general, the shot-noise component has a much higher signal-to-noise ratio than the clustering one, it cannot be used to put independent constraints on log$A$ and $\gamma$. Full power spectrum measurements are crucial to break such degeneracy, and reconstruct the \CII LF. In our fiducial survey S1 (inspired by CCAT-p/1000 hr) at $z \sim 6$, the shot-noise (clustering) signal is detectable for 2 (1) of the 5 considered $L_{\rm CII}-{\rm SFR}$ relations. The shot-noise is generally dominated by galaxies with $L_{\rm CII}\gtrsim 10^{8-9}~L_\odot$ ($M_{\rm UV}\sim-20$ to $-22$), already at reach of ALMA pointed observations. However, given the small field of view of such telescope, an IM experiment would provide unique information on the bright-end of the LF. The detection depth of an IM experiment crucially depends on the (poorly constrained) $L_{\rm CII}-{\rm SFR}$ relation in the EoR. If the $L_{\rm CII}-{\rm SFR}$ relation varies in a wide log$A$ -- $\gamma$ range, but still consistent with ALMA \CII LF upper limits, even the signal from galaxies with $L_{\rm CII}$ as faint as $\sim 10^7~L_\odot$ could be  detectable. 
Finally, we consider the contamination by continuum foregrounds (cosmic infrared background, dust, CMB) and CO interloping lines, and derive the requirements on the  residual contamination level to reliably extract the \CII signal.
\end{abstract}

\begin{keywords}
galaxies: high-redshift, dark ages, reionization, first stars, diffuse radiation, radio lines: galaxies
\end{keywords}

\section{Introduction}\label{introduction}

The intensity mapping (IM) of emission lines from galaxies provides statistical information on the properties and large scale structure distribution of these systems (for a review, see e.g. \citealt{IMreport2017}). Many candidate emission lines for IM experiments have been proposed. Among these are the \HI 21cm line (e.g. \citealt{Chang2008,Xu2015,Xu2016,Obuljen2018,Dinda2018}), the CO rotational lines with different $J$ numbers (e.g.\citealt{Lidz2011,Breysse2014,Mashian2015,Breysse2016,Breysse2017a,Breysse2017b}), the \CII  fine-structure emission line (e.g.\citealt{Gong2012,Silva2015,Uzgil2014,Yue2015,Dumitru2018,Moradinezhad2018,Padmanabhan2018}), the Ly$\alpha$ emission line (e.g. \citealt{Silva2013,Gong2014,Comaschi2016a,Comaschi2016b,Comaschi2016c}),  the HeII emission line \citep{Visbal2015}, the $^3$HeII hyper-fine spin flip transition line \citep{Bagla2009,Takeuchi2014}, and more other lines \citep{Gong2017,Fonseca2017}.

Since the IM signal is the collective radiation from galaxies with all luminosities, in principle, this technique is suitable to study faint galaxies located in the Epoch of Reionization (EoR) which cannot be individually resolved by current surveys. To isolate the faint galaxies we can mask resolved galaxies (and their PSF) found in galaxy catalogs at same/other wavelengths. This strategy has been successfully applied, e.g., to study galaxies that are unresolved in the cosmic infrared background (CIB), by masking bright galaxies using their positions from {\it HST} and {\it Spitzer} catalogs (e.g. \citealt{Kashlinsky2005,Kashlinsky2007,Kashlinsky2012,Cooray2012Nature,Zemcov2014Science}). Recent progress in observations and theories of CIB could be found in the review \citet{Kashlinsky2018}.
The remaining flux is from the targeted very faint galaxies. To avoid losing too many pixels, the IM map must have high angular resolution and smaller PSF wings. 

The \CII line, with rest-frame wavelength 157.7 $\mu$m, is generally the strongest metal line \citep{Visbal2011,Kogut2015} as it represents one of the major cooling channels of the interstellar medium.
For EoR galaxies, the line is redshifted into the sub-mm/mm bands, and hence it is at reach of ground-based radio telescopes. 
Whether the \CII line is a good star formation rate (SFR) tracer is a highly debated question, particularly at high redshifts \citep{deLooze2014,Vallini2015,Herrera-Camus2015,Zanella2018,Herrera-Camus2018,Carniani2018}. The reason is that the intensity of the line might be affected by gas metallicity, the strength of the interstellar radiation field and a number of feedback processes. Currently, the relation between  \CII line luminosity and star formation rate at high-$z$ is still not well-known, neither from theory nor observation. This problem effectively limits the precision with which the \CII IM signal from high-$z$ galaxies can be predicted.

For an IM experiment, a further challenge is represented by foreground radiation and contamination by interloping lines. For 21cm, the foreground removal has been widely investigated (e.g. \citealt{Wang2006,Liu2009,Jelic2008,Switzer2015,Zhang2016}). Foreground removal algorithms for other lines have also been presented recently \citep{Silva2015,Yue2015,Breysse2015,Lidz2016,Cheng2016,Chung2017, Sun2018}. In the light of these progresses, it is conceivable that robust solutions to the foreground/contamination removal problem will be available in the near future. Assuming this hypothesis, in this paper we base our analysis on a foreground/contamination free \CII IM signal. In Sec. \ref{sec:foreground} we discuss  the level of continuum foregrounds and contaminating lines, and test the feasibility of two cleaning methods

Cross-correlating different signals strongly reduces the impact of foregrounds and contaminating lines. These experiments may well lead to an earlier discovery of cosmological \CII emission with respect to those searching for the auto-correlation signal. Cross-correlations between IM signals of different emission lines (e.g. \citealt{Chang2015,Serra2016}) and between IM signals and galaxies  have been proposed.  For example the Ly$\alpha$ IM and Lyman Alpha Emitters (LAEs) cross-correlation \citep{Comaschi2016b}; the Ly$\alpha$ IM and \CII IM cross-correlation \citep{Comaschi2016c}; the \HI 21cm - galaxy cross-correlation
\citep{Furlanetto2007,Lidz2009}; the 21cm-LAEs cross-correlation \citep{Vrbanec2016,  Sobacchi2016,Feng2017,Hutter2018,Kubota2018,Yoshiura2018};
the near-infrared background (NIRB) - \HI 21cm cross-correlation \citep{Fernandez2014,Mao2014}; the NIRB - galaxy cross-correlation \citep{Yue2016}; the \CII and \HI 21cm cross-correlation \citep{Gong2012}. Recently, by cross-correlating the {\it Planck} 545 GHz data with the quasars and CMASS galaxies in the SDSS-III survey, \citet{Pullen2018}  found an anomalous signal that could be interpreted as \CII emission, albeit alternative explanations cannot be excluded.

In the following we will  predict the capabilities of $z\sim6$ \CII~IM experiments\footnote{Throughout the paper,  we assume a flat Universe with the following cosmological parameters:  $\Omega_{\rm m} = 0.308$, $\Omega_{\Lambda} = 1- \Omega_{\rm m} = 0.692$, $\Omega_{\rm b} = 0.048$, $h=0.678$, $\sigma_8=0.815$ and $n_s=0.968$, where $\Omega_{\rm m,}$, $\Omega_{\Lambda}$, $\Omega_{\rm b}$ are the total matter, vacuum, and baryonic densities, in units of the critical density,   and  $h$ is the Hubble constant in units of 100 km/s.
$\sigma_8$ is the normalized density fluctuations and $n_s$ is the index of the primordial fluctuations \citep{Planck2015cos}. 
}. 
After some preliminary considerations, our analysis proceeds through the following steps.
We start by deriving the \CII luminosity function (LF) from the observed UV LF, followed by the IM signal. We predict the amplitude of the power spectrum and its detectability, for shot-noise and full power spectrum, respectively. Shot-noise is the regime where the first observational results are likely to become available. We also quantify the fraction of the IM signal that comes from galaxies with different \CII luminosities and investigate the feasibility of using IM to detect galaxies fainter than those observed by pointed observations. The final step is to investigate the possibility to recover the \CII LF from the IM observation, and the \CII luminosity density using a range of survey strategies.

\section{Preliminary considerations}

\subsection{\CII IM experiments and sensitivities}\label{telescopes}

Several instruments have been proposed -- or are in the development phase, to conduct \CII IM experiments. These include the 6 m CCAT-p \citep{Parshley2018a,Parshley2018b,Vavagiakis2018}\footnote{\url{www.ccatobservatory.org/}}, the 12 m CONCERTO  \citep{Serra2016,Dumitru2018}\footnote{\url{people.lam.fr/lagache.guilaine/CONCERTO.html}}, and the TIME-pilot \citep{Crites2014} to be installed on the JCMT\footnote{\url{www.eaobservatory.org/jcmt/}}. Generally, these instruments will typically feature spectrometers with resolving power $R=\nu_0/\delta \nu_0 \sim 100-300$ and thousands cryogenic detectors.

The CCAT-p would have noise-equivalent flux density ${\rm NEFD}=2.5\times10^{6}$ Jy s$^{1/2}$sr$^{-1}\Omega_{\rm beam}$ for $\delta \nu_0=0.4$ GHz \citep{Stacey2018,Padmanabhan2018}. Since the beam angular size is
\begin{equation}
\Omega_{\rm beam}\sim\left(1.22 \frac{\lambda_0}{D}\right)^2,
\end{equation}
and 
\begin{align}
{\rm NEFD}&\sim\frac{2 k_{\rm B} T_{\rm sys}}{D^2\sqrt{\delta \nu_0} },
\end{align}
where $k_{\rm B}$ is the Boltzmann constant, this would correspond to a system temperature $T_{\rm sys}\sim 33$ K at frequency $\nu_0=272$ GHz. 
CONCERTO will have ${\rm NEFD}= 155$ mJy s$^{1/2}$ \citep{Dumitru2018} for $\delta \nu_0=1.5$ GHz, corresponding to $T_{\rm sys}\sim246$ K. TIME-pilot \citep{Crites2014} will work at the background-limited noise level, with NEFD $\sim60-70$ mJy s$^{1/2}$ in the high-frequency band.

\subsection{Pointed vs. intensity mapping observations}

Before diving into a detailed description of our IM model, we start by examining a more general question about the relative advantages of pointed and IM observations for investigating distant galaxies. In particular, we focus on a specific question: can high-$z$ \CII~IM experiments detect the signal from galaxies too faint to be detected by pointed observations? To answer this question we consider a practical case involving the largest spectroscopic survey carried out so far by ALMA in the Hubble Ultra Deep Field (UDF, \citealt{Aravena2016}),  and compare its findings with those from a recent CO IM experiment, COPSS I \& II \citep{Keating2015COPSSI,Keating2016}. 

The ALMA UDF survey \citep{Walter2016} performs a spectral scan of a sky region in the UDF field with total mosaic area of $\sim 1$ arcmin$^2$. The Band 6 \citep{Aravena2016} scan covers the frequency range $212-272$ GHz with steps of 31.25 MHz per channel, suitable to detect the \CII~line in $6\lsim z\lsim 8$. The total comoving volume of the survey is 4431 Mpc$^3$. The field has been observed for about 20 hr with a synthesized beam of 1.5\arcsec  $\times$ 1.0\arcsec. The average r.m.s. noise level is $\sigma_{\rm N} = 0.53$ mJy beam$^{-1}$. Assuming a line width $\Delta v =300\, \rm km~s^{-1}$, the 3$\sigma$ detection limit in term of the \CII~luminosity is $(1.6-2.5) \times 10^8~L_\odot$ in the survey redshift range. In the surveyed UDF volume, {\it HST} found 58 galaxies at $z>5.5$; {ALMA} found 14 \CII emitters at $> 4.5\sigma$ level, 8 of which are considered as statistically spurious; 2 out of 14 are blind detections with no optical/IR counterparts.  In addition to Band 6, the same UDF sky region has been observed also in ALMA Band 3 ($84 -115$ GHz, \citealt{Decarli2016}) with spectral resolution of 19.5 MHz. In this band, 7 objects with detected CO(3-2) line ($2.01 < z <3.02$) were found in the corresponding survey volume 3363 Mpc$^3$.
\begin{table*}
\caption{Comparison between the main features of the ALMA UDF \citep{Walter2016,Aravena2016,Decarli2016} observations and the COPSS IM experiments \citep{Keating2015COPSSI,Keating2016}.}
\begin{tabular}{|c|c|c|c|c|c|c|c|c|c|c|c|c|c|}
\hline\hline
 Survey       &&ALMA Band6 & ALMA Band3 & COPSS I & COPSS II \\ 
Line&---&\CII&CO(3-2)&CO(1-0) & CO(1-0)\\ 
Band & [GHz]&$212-272$ &$84-115$ & $27-35$ & $27-35$\\
Spectral resolution & [GHz] &0.0313 &0.0195 &0.0313 &0.0313 \\
Redshift &---&$6-8$&$2.01-3.02$&$2.3-3.3$ & 2.3-3.3\\
Beam & ---&1.5\arcsec$\times$1.0\arcsec&3.5\arcsec$\times$2.5\arcsec&$\sim(0.5\arcmin)^2$& $\sim(2\arcmin)^2$\\
Survey area &--- &75\arcsec$\times$70\arcsec&90\arcsec$\times$90\arcsec&1.7deg$^2$&0.7deg$^2$\\
Survey volume & [$h^{-3}$Mpc$^3$]&1520&3363&$3.6\times10^6$ & $1.3\times10^6$\\
Observational time &[hr]&20&20&900&3000\\
Instrumental noise &[mJy beam$^{-1}$] &0.5&$0.15$&0.15&0.05\\
---&[Jy sr$^{-1}$]&$5.4 \times 10^6$&$7 \times 10^5$&$\sim7\times10^3$&$\sim150$\\
\hline \\
\end{tabular}\\
\label{Table1}
\end{table*}

The COPSS I/II project \citep{Keating2015COPSSI,Keating2016} performs an IM experiment aimed at detecting the CO(1-0) transition at 115.2712 GHz using the SZA array in the frequency range $27-35$ GHz (corresponding to $2.3 < z < 3.3$) over 44/19 fields, covering an area of 1.7/0.7 square degrees, thus resulting in a total surveyed volume of $4.9\times 10^6 h^{-3}\rm Mpc^3$. The total observation time 
for the 19 COPSS II fields is of about 3000 hr, resulting in an average r.m.s. noise level  $\sigma_{\rm N} = 0.05$ mJy beam$^{-1}$. The experiment has been able to measure the amplitude of the power spectrum,
\be 
\Delta^2_{\rm CO} (k= 1~h\, \textrm{Mpc}^{-1}) = \frac{k^3}{2\pi^2} P_{\rm CO} =  1.5 ^{+0.7}_{-0.7} \times 10^3 \mu\rm K^2, 
\ee
 corresponding to a fluctuation amplitude $2(\nu_0/c)^2 k_{\rm B}\Delta_{\rm CO} \approx 1.1\times10^3\, \rm Jy\, sr^{-1}$. 
 
The parameters of the {ALMA} UDF survey and the COPSS IM experiment are summarized in Tab. \ref{Table1}. By comparing them we can appreciate the main differences between pointed and IM observations. 
The COPSS I IM experiment covers a volume $\sim1000\times$ larger than the ALMA Band 3 survey in an observing time that is only 
$\sim50$ times longer, reaching at the same time similar noise. Note also {ALMA}'s much worse brightness sensitivity compared with the SZA ($7\times 10^5\, \rm Jy\, sr^{-1}$ vs. $7\times10^3\, \rm Jy\, sr^{-1}$): this reflects the well known disadvantage of interferometers with respect to compact dishes in detecting extended emission due to their small beam. Of course, IM experiments must yield in terms of angular resolution which, in this case, is $\gtrsim$10 times worse. 

Generally speaking, the competition between pointed and IM observations depends on the details of the emission line LF. For galaxies with $0.5 < z < 1.5$ this issue has been investigated by \citet{Uzgil2014}.
They obtained the emission line LF from the observationally constrained IR LFs and the line luminosity - IR luminosity relations.
Their generic conclusion is that while wide-field galaxy surveys are not deep enough, deep pencil beam surveys do not collect large enough galaxy samples. As a result, it appears that IM is the best way to observe the total emission from galactic populations. 

\section{Method}\label{Met}
 
\subsection{The \CII luminosity function}\label{CII_LF}
 
To derive the \CII LF, we start from the galaxy UV LF \citep{Schechter1976}, 
\begin{align}
\frac{dn}{dM^{\rm obs}_{\rm UV}}= & \, 0.4\,{\rm ln}(10)\, \Phi^*  10^{-0.4(M^{\rm obs}_{\rm UV}-M_{\rm UV}^*)(\alpha+1)}           \nonumber \\
& \times  {\rm exp}[-10^{-0.4(M^{\rm obs}_{\rm UV}-M_{\rm UV}^*)}],
\label{UVLF}
\end{align}
where $M^{\rm obs}_{\rm UV}$ is the observed (dust-obscured) absolute UV magnitude.
At $z\sim6$, $\Phi^*=(0.50^{+0.22}_{-0.16})\times10^{-3}$ Mpc$^{-3}$, $M_{\rm UV}^*=-20.94\pm0.20$, $\alpha=-1.87\pm0.10$, constrained  by 857 galaxies in {\it HST} datasets \citep{Bouwens2015}.

The dust-corrected absolute UV magnitude is  $M_{\rm UV}=M^{\rm obs}_{\rm UV}-A_{\rm UV}$, where the dust attenuation $A_{\rm UV}$ can be derived from the infrared excess -- UV spectral slope relation (IRX-$\beta$ relation). It is found that $A_{\rm UV}$ depends linearly on the UV spectrum slope $\beta$ \citep{Meurer1999},
\begin{equation}
A_{\rm UV}=C_0\beta+C_1~~~~~~~~(A_{\rm UV}\ge 0),
\label{IRX}
\end{equation}
and $\beta$ could be derived from the observed UV magnitude 
\begin{equation}
\beta=\beta_0 + \frac{d\beta}{dM_0}(M^{\rm obs}_{\rm UV}-M_0).
\end{equation}
At $z\sim6$ \citet{Bouwens2014a} found  $\beta_0=-2.00\pm0.05\pm0.08$, $d\beta/dM_0=-0.20\pm0.04$ and $M_0=-19.5$ from 211 Lyman Break Galaxies (LBGs) observed by {\it HST}.
Then the UV LF for the dust-corrected UV magnitude $M_{\rm UV}$ is simply 
\begin{equation}
\frac{dn}{dM_{\rm UV}}=  \frac{dn}{dM^{\rm obs}_{\rm UV}}  \frac{dM^{\rm obs}_{\rm UV}}{dM_{\rm UV}}.
\end{equation}

Since the original \citet{Meurer1999} proposal, many works have re-derived  the coefficients $C_0$ and $C_1$ from the IRX-$\beta$ relation. In general, these studies \citep{Buat2012,Talia2015,Forrest2016,Bourne2017,McLure2018,Koprowski2018} reach similar conclusions. However, some discrepancies remain, as pointed out by e.g. \citealt{Talia2015}, and \citet{Reddy2018}. These differences may arise from a sample selection bias or the different algorithms used to determine  $\beta$.  

There is also a non-negligible scatter in the $A_{\rm UV}-\beta$ relation. \citet{Koprowski2018} found a scatter $\lesssim 0.2$ dex; \citet{Talia2015} claimed a larger scatter ($\sim 0.7$ dex) for their mass-selected galaxies with wide star formation histories and different starburst activity levels. On the other hand, \citet{Salim2019} pointed out that 
the errors on the SFR derived from the IRX-$\beta$ relation should be $\lesssim$ 0.15 dex, corresponding to a dust attenuation uncertainty  $\lesssim0.4$ dex.

In this paper we adopt $C_0=2.10$ and $C_1=4.85$, as obtained by \citet{Koprowski2018} from 4178 LBGs with $z=3-5$. Although \citet{Koprowski2018} quote a scatter $\lesssim0.2$ dex, we conservatively use in Eq. (\ref{IRX}) a larger value (0.4 dex r.m.s.), inspired by the larger scatters found by other studies. This choice also makes our assumptions safely consistent with the original \citet{Meurer1999} $C_0=1.99$ and $C_1=4.43$ values.

\begin{figure}
\includegraphics[width=80mm]{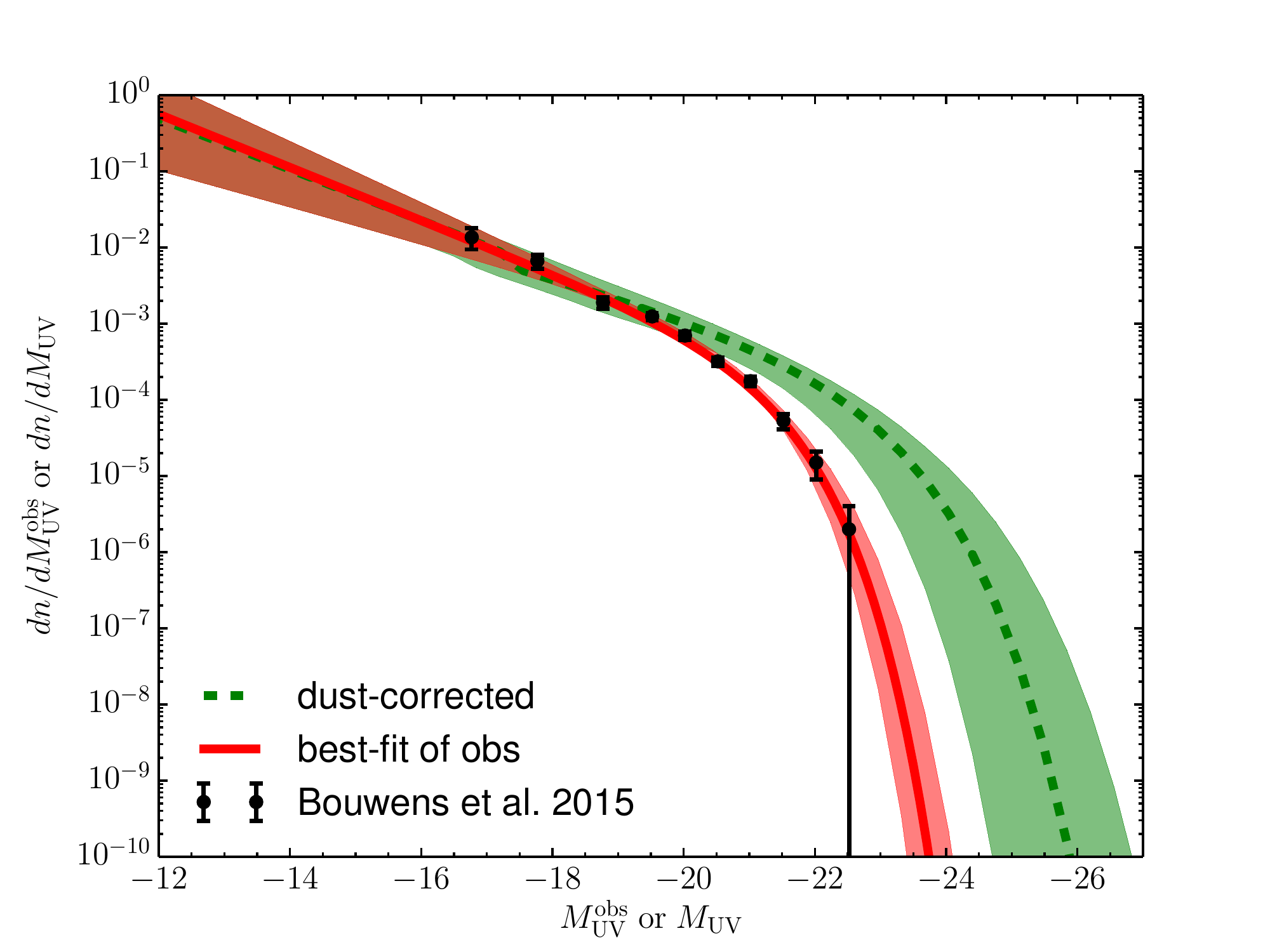}
\caption{
The observed (red, solid) and dust-corrected (green, dashed) UV LFs at $z\sim6$, shaded regions refer to $1\sigma$ uncertainties. 
}
\label{fig:UVLF}
\end{figure}

In Fig. \ref{fig:UVLF} we show the UV LFs vs. observed and dust-corrected magnitudes, along with their $1\sigma$ uncertainties.  The latter arise from the observed UV LF, the observed $\beta-M^{\rm obs}_{\rm UV}$ relation, and the IRX-$\beta$ relation.

\subsection{The \CII -SFR relation and its uncertainties}\label{CII_SFR}

Most of the physics of the problem is contained in the \CII luminosity ($L_{\rm CII}$) dependence on the SFR of a galaxy. Broadly speaking, such relation arises from a balance between stellar energy production rate, and the ability of the gas to radiate away this energy via one of the most powerful coolants, i.e. the \CII line. Modelling the underlying physics in detail is difficult, albeit the sophistication level of some recent studies is increasing. Therefore, one resorts to local empirical determinations, and extrapolates them to higher redshifts. 

A commonly used parameterisation of the  $L_{\rm CII}-{\rm SFR}$ (with SFR in units of $M_\odot {\rm yr}^{-1}$ and $L_{\rm CII}$ in units $L_\odot$) relation is a power-law form plus a scatter
\begin{align}
{\rm log}(L_{\rm CII})&={\rm log}(\bar{L}_{\rm CII})\pm \sigma_L \nonumber \\
&={\rm log}A+\gamma {\rm log}({\rm SFR})\pm\sigma_L,
\end{align}
where SFR is proportional to the intrinsic UV luminosity \citep{Kennicutt1998}
\begin{equation}
 {\rm SFR}={\mathcal K}_{\rm UV}L_{\rm UV}, 
\end{equation}
and ${\mathcal K}_{\rm UV}\approx0.7\times10^{-28} M_\odot {\rm yr}^{-1}/{\rm erg~s^{-1}Hz^{-1}}  $ \citep{bc03} for a
\citet{Chabrier2003} stellar initial mass function (IMF),  metallicity in the range 0.005-0.4$Z_\odot$, and stellar age $\gtrsim100$ Myr. The factor ${\mathcal K}_{\rm UV}$ depends somewhat on galaxy metallicities and stellar ages.
We have checked that ${\mathcal K}_{\rm UV}$ increases only to $\approx0.9\times10^{-28} M_\odot {\rm yr}^{-1}/{\rm erg~s^{-1}Hz^{-1}}$ for a metallicity as high as 2.5$Z_\odot$. Also, at $z\sim6$, except for very young starburst systems, the stellar ages of most galaxies are likely $ > 100$ Myr \citep{Salvaterra11, Hashimoto2018}. Therefore,  the scatter on ${\mathcal K}_{\rm UV}$ should have a small influence on our results; we ignore it in the rest of the paper.

\begin{figure}
\includegraphics[width=90mm]{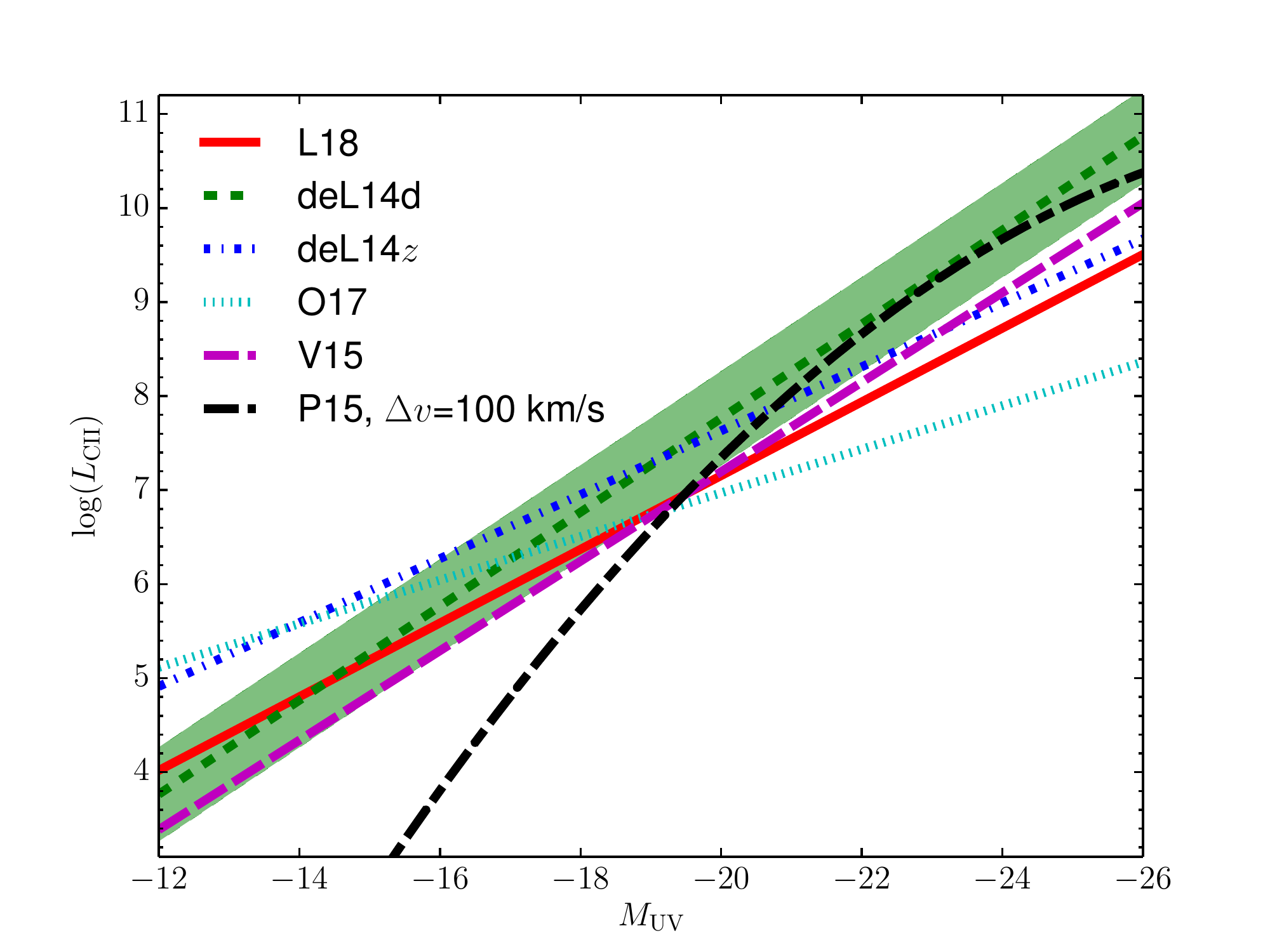}
\caption{The ${\rm log}(L_{\rm CII})-M_{\rm UV}$ relation at $z\sim 6$ deduced from various observations and simulations (a)-(e) and simulations (P15) described in the text. To avoiding the panel we only plot the scatters for deL14d case.
}
\label{fig:logLCII-MUV}
\end{figure}

Several works have studied the $L_{\rm CII}-{\rm SFR}$ relation, both theoretically (using numerical simulations and/or post-processing techniques, and semi-analytical models) and observationally. 
\citet{Vallini2015} (V15) used hydrodynamical simulations coupled with a sub-grid ISM model to compute the \CII emission from diffuse neutral gas and photodissociation regions, for  galaxies with varying SFR and metallicity. Their results are fitted by \begin{align}
{\rm log}(\bar{L}_{\rm CII})& = 7.0 + 1.2\times {\rm log(SFR)} + 0.021\times {\rm log}(Z)\nonumber \\
&+ 0.012\times{\rm log(SFR)log(Z)}-0.74\times{\rm log}^2(Z).
\end{align}
For $Z=0.2Z_\odot$, log$A=6.62$ and $\gamma=1.19$. Variance in metallicity leads a corresponding scatter in \CII luminosity. For example, 
if the scatter on $Z$ is $\sim$0.4 dex \citep{Salvaterra11}, \CII luminosities vary by about $\sim0.4$ dex for SFR $=10~M_\odot {\rm yr^{-1}}$.

\citet{Olsen2017} (O17) combined cosmological galaxy formation simulations (a re-simulation technique was used) with {\tt S\'IGAME} \citep{Olsen2015} to investigate the \CII, \OI and \OIII emission from 30 selected $z\sim6$ galaxies in a relatively narrow metallicity ($0.1-0.4 Z_\odot$), SFR ($3-23 M_\odot$yr$^{-1}$) and stellar mass ($0.7\times10^9$-$8\times10^9~M_\odot$) range. They found that most \CII emission arises from diffuse ionized gas and molecular clouds. 
They obtain log$A=6.69\pm0.10$ and $\gamma=0.58\pm0.11$, with a \CII scatter $\sim 0.15$ dex; note however, that this estimate is based only on the 30 selected galaxies. 

\citet{Lagache2018} (L18) used a semi-analytical galaxy evolution model {\tt GAS}, together with the photoionization code {\tt CLOUDY} to determine the galaxy properties in a box with size $100$ $h^{-1}$Mpc$^{-1}$ and derive  the \CII line luminosity for a large number of galaxies at $4<z<8$, assuming that \CII emission is mostly from photodissociation regions. 
Their best-fit $L_{\rm CII} -{\rm SFR}$ relation is
\begin{align}
{\rm log}A&=7.1-0.07z \nonumber \\
\gamma&=1.4-0.07z;
\end{align}
e.g., at $z=6$ log$A=6.68$ and $\gamma=0.98$, with a scatter of 0.60 dex.  Their predicted \CII luminosity is smaller than observed for local dwarf galaxies \citep{deLooze2014}. Such discrepancy cannot be explained by metallicity or CMB attenuation effects. They conclude that the \CII scatter arises from variations of interstellar radiation field, neutral gas density and metallicity. 

\citet{deLooze2014} analyzed the relation between \CII emission and SFR for various galaxy populations, including local metal-poor dwarf galaxies and high-$z$ samples. The local dwarf galaxies sample, which is composed of 42 galaxies, is mostly selected from the Dwarf Galaxy Survey (DGS) carried by {\it Herschel}. Galaxies in this sample feature star formation rates  (derived from UV+IR data) in the range ${\rm SFR}\sim 0.001-43 M_\odot {\rm yr}^{-1}$ and metallicities in the range $0.03-0.55~Z_\odot$. For these dwarf galaxies, they found log$A=7.16$ and $\gamma=1.25$, with a scatter 0.5 dex. The high-$z$ sample has 27 galaxies, with redshift range $z=0.59-6.60$ and SFR$=69.2-67,000~M_\odot{\rm yr}^{-1}$. The SFR in this sample is estimated from the total infrared luminosity. They found log$A=7.22$ and $\gamma=0.85$, with a scatter $\sim0.3$.

In summary, in the following we will adopt the above five  $L_{\rm CII} -{\rm SFR}$ relations to investigate the \CII IM signal:{\bf(a)}   \citet{Lagache2018} (L18, ${\rm log}A=6.68,\gamma=0.98$, scatter 0.60 dex),
{\bf(b)} \citet{deLooze2014} metal-poor dwarf galaxies at $z\sim0$ (deL14d, ${\rm log}A=7.16,\gamma=1.25$, scatter 0.5 dex), 
{\bf (c)} \citet{deLooze2014} high-$z$ samples (deL14$z$, ${\rm log}A=7.22,\gamma=0.85$, scatter 0.3 dex),
{\bf (d)}  \citet{Olsen2017} (O17, ${\rm log}A=6.69,\gamma=0.58$, scatter $\sim$0.15 dex) and {\bf (e)} \citet{Vallini2015}  for $0.2Z_\odot$ metallicity (V15, ${\rm log}A=6.62,\gamma=1.19$, scatter $\sim$0.4 dex). Note that for V15 the scatter estimate is for  a metallicity scatter of $\sim0.4$ dex when SFR=10$M_\odot {\rm yr}^{-1}$. They are listed in Tab. \ref{tab:LCII_SFR}.

 \begin{table*}
 \centering{
 \caption{Summary of $L_{\rm CII} -{\rm SFR}$ relation parameters we adopt.} 
\begin{tabular}{ccccccccc} 
\hline 
  & log$A$ & $\gamma$ & $\sigma_L$&$z$&number of samples&references \\
\hline
L18 & 6.68 & 0.98 & 0.60& $\sim6$& all galaxies in 100 $h^{-1}$Mpc box&\citet{Lagache2018}\\
deL14d & 7.16 & 1.25 &0.5& local&42&\citet{deLooze2014}\\
deL14$z$ & 7.22 & 0.85 &0.3& 0.59-6.60& 27&\citet{deLooze2014}\\
O17 &6.69& 0.58 & 0.15&6&30&\citet{Olsen2017}\\
V15 & 6.62 & 1.19 &0.4&$\sim6.6$&one however varying $Z$ and SFR& \citet{Vallini2015} \\
\hline
\end{tabular}
\label{tab:LCII_SFR}
}
\end{table*}

Additional studies have investigated the magnitude of the scatter in the relation.  \citet{Herrera-Camus2015} found a  $\sim$0.2 dex scatter; \citet{Carniani2018} obtained $0.48$ dex; in \citet{Herrera-Camus2018} for star-forming galaxies the scatter is 0.32 dex. In principle, the scatter from observations includes both observational uncertainties and intrinsic scatter; it may also depend on how the sample was selected. Scatter in simulations may suffer from small sample size and/or the algorithms used to model the line radiative process. These caveats must be kept in mind and the robustness of the reported scatter should not be over-emphasized as constraints are still poor. 

Further constraints come from the \CII LFs. In order not to exceed the upper limits on $z\sim6$ CII LF (e.g.  \citealt{Yamaguchi2017}), the scatter cannot exceed the following values:  for L18, $\sigma_L\lesssim1.3$ dex, for deL14d, $\sigma_L\lesssim1.1$ dex, for deL14$z$, $\sigma_L\lesssim1.1$ dex, for O17 $\sigma_L\lesssim1.2$ dex, for V14 $\sigma_L\lesssim1.2$ dex. However, these should be considered only as extreme upper limit constraints. Note that the scatter may also depend on SFR. 
To avoid complicating the model, in following we adopt the reported scatter of each relation; for other values of log$A$ and $\gamma$ we used instead a fiducial scatter of 0.4 dex.
  
In all simulations the SFR entering the $L_{\rm CII}-{\rm SFR}$ relation refers to the intrinsic SFR.  However,  \citet{deLooze2014} observations use a SFR value derived by combining UV and IR data.  The consistency between these two methods has been checked. In our paper, the SFR is from the dust-corrected UV and, when correcting for dust-attenuation, the IR data is used for calibration. By definition these estimates should be all consistent with each other\footnote{In principle even when the SFR is specified, the \CII emission still depends on the stellar IMF. Both L18 and O17 used the \citet{Chabrier2003} IMF.  \citet{deLooze2014} used the \citet{Kroupa2003} IMF for which ${\mathcal K}_{\rm UV}$ is close to that of the \citet{Chabrier2003} IMF. V15 specified a Salpeter IMF between $1-100~M_\odot$, leading to similar ${\mathcal K}_{\rm UV}$, therefore in this paper no IMF correction for SFR is applied.}; in practice, this might not be strictly true as a result of uncertainties in IR SEDs, stellar IMF, and the algorithms to model the CII emission process.

Also, different galaxy populations (starburst, normal star-forming galaxies, metal-poor dwarf galaxies, etc) may have different 
$L_{\rm CII}-{\rm SFR}$ relations and the relations may evolve with redshift. Therefore if samples are selected by galaxy population or redshift, the derived $L_{\rm CII}-{\rm SFR}$ relation may vary. Moreover, if only \CII-detected galaxies are selected, then the relation could be biased towards the most luminous \CII emitters. Errors in the SFR derivation, and limited sample sizes would also bias the derived $L_{\rm CII}-{\rm SFR}$ relations. At present, these factors are not under full control, particularly for the relatively few high-$z$ \CII-detected  galaxies. To bracket all these uncertainties, we limit ourselves to present \CII IM predictions for all the five $L_{\rm CII}-{\rm SFR}$ relations discussed above.

Replacing SFR with $M_{\rm UV}$, we have 
\begin{equation}
{\rm log}\bar{L}_{\rm CII}={\rm log}A-0.4\, \gamma\,(M_{\rm UV}+18.79).
\label{eq:logLCII_MUV}
\end{equation}
The considered relations are shown in Fig. \ref{fig:logLCII-MUV}. 
To clearly show all the lines we only present the deL14d scatter.
 In practice, we are dealing with galaxies brighter than $M_{\rm UV}\sim-20$ to $-22$ if the \CII luminosity is larger than $\sim10^8~L_\odot$. 
As a comparison, in the same panel we also show the relation obtained in \citet{Pallottini2015} (P15). The P15 peak flux, $F$, is converted into a \CII~line luminosity through 
\begin{equation}
L_{\rm CII}=F_{\rm CII}\times\left( \frac{\Delta v}{1+z}{\nu_\mathrm{CII}\over c}\right) \times4\pi d_L^2,
\label{eq_flux}
\end{equation}
assuming a linewidth $\Delta v = 100 \kms$. For sake of simplicity, we will not use the P15 relation as its functional form is not well described by the simple power-law Eq. (\ref{eq:logLCII_MUV}).

Considering the log-normal distribution of  $L_{\rm CII}$ with scatter $\sigma_L$,
the probability to find a galaxy with magnitude $M_{\rm UV}$ and luminosity $L_{\rm CII}$ is 
\begin{align}
&P({\rm log}L_{\rm CII}|M_{\rm UV})d{\rm log}L_{\rm CII}= \nonumber \\
&\frac{1}{\sqrt{2\pi}\sigma_{L }}{\rm exp}\left[- \frac{  (  {\rm log}L_{\rm CII}-{\rm log}\bar{L}_{\rm CII}  )^2 }{2\sigma^2_{L}  }  \right]d{\rm log}L_{\rm CII}.
\end{align}
The final step to obtain the \CII LF requires an integration over the UV magnitude,
\begin{equation}
\frac{dn}{d{\rm log}L_{\rm CII}}=\int dM_{\rm UV}\frac{dn}{dM_{\rm UV}}    P({\rm log}L_{\rm CII}|M_{\rm UV}).
\label{dndL}
\end{equation}

\begin{figure}
\includegraphics[width=90mm]{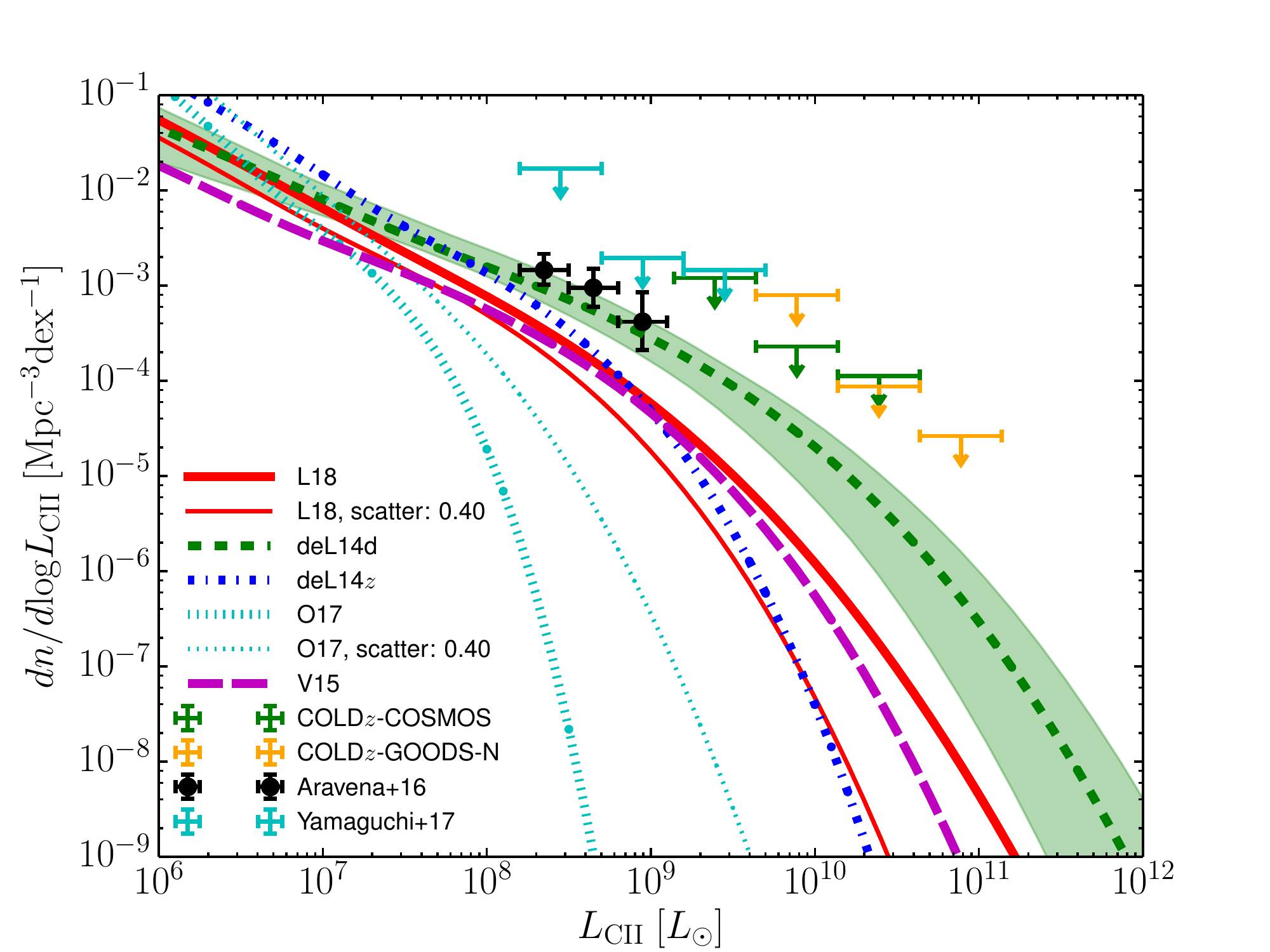}
\caption{The $z\sim6$ \CII LFs derived from the UV LF for different $L_{\rm CII} - {\rm SFR}$ relations listed in the legend, compared to the observational data/upper limits by \citet{Aravena2016} and \citet{Yamaguchi2017}. Also shown are the upper limits derived from the CO LFs at $z \approx 5.8$ in the COLDz survey \citep{Riechers19}, by assuming $L_{\rm CII}/L_{\rm CO(1-0)}\sim4400$  and $T_{\rm CO(2-1)}/T_{\rm CO(1-0)} \sim 1$. Note that \citet{Aravena2016}  actually give $n(>{\rm log}L_{\rm CII})$; in the plot we use approximation $dn/d{\rm log}L_{\rm CII}({\rm log}L_{\rm CII})\sim n(>{\rm log}L_{\rm CII})$ which is valid when LF drops fast with increasing ${\rm log}L_{\rm CII}$. For clarity,  we only show the uncertainties for the deL14d case.
}
\label{fig:CIILF}
\end{figure}

We plot the \CII LFs at $z\sim6$ in Fig. \ref{fig:CIILF} for the  $L_{\rm CII} - {\rm SFR}$ relations ({\bf a})-({\bf e}) introduced above (thick lines). For comparison, for model L18 (O17) which has largest (smallest) scatter, we also plot the \CII LF for an assumed to scatter of 0.4 dex, i.e., in between the two extreme values (thin lines).
In the same panel we plot the data\footnote{The \citet{Aravena2016} data should be considered as upper limits, as their sources are candidates.} from \citet{Aravena2016}, the
upper limits from \citet{Yamaguchi2017}, and those derived from the CO LFs at $z \approx 5.8$ in the COLDz survey \citep{Riechers19}, by assuming $L_{\rm CII}/L_{\rm CO(1-0)}\sim4400$ (e.g. \citealt{Stacey2010,Swinbank2012}) and $T_{\rm CO(2-1)}/T_{\rm CO(1-0)} \sim 1$ \citep{Riechers19}. 

\subsection{Intensity mapping}

The observed IM signal is expected show spatial fluctuations due to: (i) clustering (CL) of large scale structures; and (ii) shot-noise (SN) due to Poisson fluctuations of the number of galaxies in a given volume around the mean value. The CL  signal  (in Fourier space) depends on wavenumber, $k$, while the shot-noise component is constant. Typically the shot-noise dominates the CL term for large $k$-values (small scales). The \CII power spectrum is the sum of the clustering and shot-noise terms,
\begin{equation}
P_{\rm CII}(k,z)=P^{\rm CL}_{\rm CII}(k,z)+P^{\rm SN}_{\rm CII}(z).
\end{equation}

The clustering  term can be written as \citep{Shang2012}
\begin{align}
P^{\rm CL}_{\rm CII}(k,z)&= \left( \frac{c}{4\pi \nu_{\rm CII}H(z)}\right)^2  \nonumber \\
&\times \left(  \int dM_h \frac{dn}{dM_h}    L_{\rm CII}(M_h)  b_{\rm SMT}  \right)^2  P(k,z),
\end{align}
where  $dn/dM_h$ is the halo mass function, $P(k,z)$ is the matter power spectrum, $b_{\rm SMT}$ is the bias for halos with mass $M_h$ taken from \citet{ST01}; the $L_{\rm CII}-M_{\rm h}$ relation is derived from abundance matching following \citet{Yue2015}.

The shot-noise is computed as
\begin{align}
P^{\rm SN}_{\rm CII}(z) &= \left({c\over 4\pi \nu_\mathrm{CII} H(z)}\right)^2\int  d{\rm log}L_{\rm CII}  {dn\over d{\rm log}L_\mathrm{CII}} L_\mathrm{CII}^2.
\label{SN}
\end{align}
The power spectrum has a variance
\begin{equation}
\sigma^2_{\rm CII}(k,z)=\frac{1}{N_m(k)} (P_{\rm CII}(k,z)+P_{\rm N})^2,
\label{sigma2}
\end{equation}
where $N_m(k)$ represents the number of sampled $k$ modes,
\begin{equation}
N_m(k) \approx{2\pi k^3d{\rm ln}k}\frac{V_{\rm survey}}{(2\pi)^3},
\end{equation}
where $d{\rm ln}k$ is the relative width of the selected $k$-bin, and $V_{\rm survey}$ is the survey volume; $P_{\rm N}$ is the instrumental noise power spectrum. The signal-to-noise ratio in $k$-bin 
\begin{equation}
{\rm S/N}= \sqrt{N_m(k)}\frac{P_{\rm CII}(k,z)}{P_{\rm CII}(k,z)+P_{\rm N}},
\label{eq:SNR}
\end{equation}
and the total signal-to-noise ratio
\begin{equation}
{\rm (S/N)_{tot}}=\left(\sum_i \frac{P^2_{\rm CII}(k)}{\sigma^2_{\rm CII}(k)}\right)^{1/2},
\end{equation}
is obtained by summing over all  $k$ bins.

Suppose the survey has angular area $\Omega_{\rm survey}$ and a frequency range [$\nu_0$,  
$\nu_0+W_0$]. Then, the survey volume is
\begin{equation}
V_{\rm survey}\approx r^2\Omega_{\rm survey} \left[  \frac{c}{H(z)} \frac{W_0 (1+z)}{\nu_0} \right].
\end{equation}
where $z=\nu_{\rm CII}/\nu_0-1$, and $r$ is the comoving distance to the redshift $z$. If the telescope has diameter $D$, system temperature $T_{\rm sys}$, and the spectrometer has  resolution $\delta \nu_0$, then the instrumental noise power spectrum is
\begin{align}
P_{\rm N}=\left[\frac{2k_BT_{\rm sys}}{D^2\sqrt{\delta \nu_0 t_{\rm vox}}}\frac{1}{\Omega_{\rm beam}}\right]^2V_{\rm vox},
\end{align}
where the comoving volume of the real space voxel is
 \begin{equation}
 V_{\rm vox}\approx r^2\Omega_{\rm beam}  \left[  \frac{c}{H(z)} \frac{\delta \nu_0 (1+z)}{\nu_0} \right].
 \end{equation}
If the detector has $N_{\rm det}$ bolometers, and the total observation time is $t_{\rm obs}$, then the integration time per voxel is 
\begin{equation}
t_{\rm vox}=\frac{t_{\rm obs} }  {(V_{\rm survey}/V_{\rm vox}/N_{\rm det}) }.  
\end{equation} 
The survey detects $k$ modes smaller than
\begin{align}
k_{\rm max}&=\sqrt{k^2_{\rm max,\perp} + k^2_{\rm max,\parallel}} \nonumber \\
&\approx\pi\sqrt{\frac{1}{r^2\Omega_{\rm beam}} +\left(\frac{1}{cH(z)^{-1}\delta \nu_0(1+z)/\nu_0}\right)^2 }
\end{align}

\begin{figure}
\subfigure{\includegraphics[width=90mm]{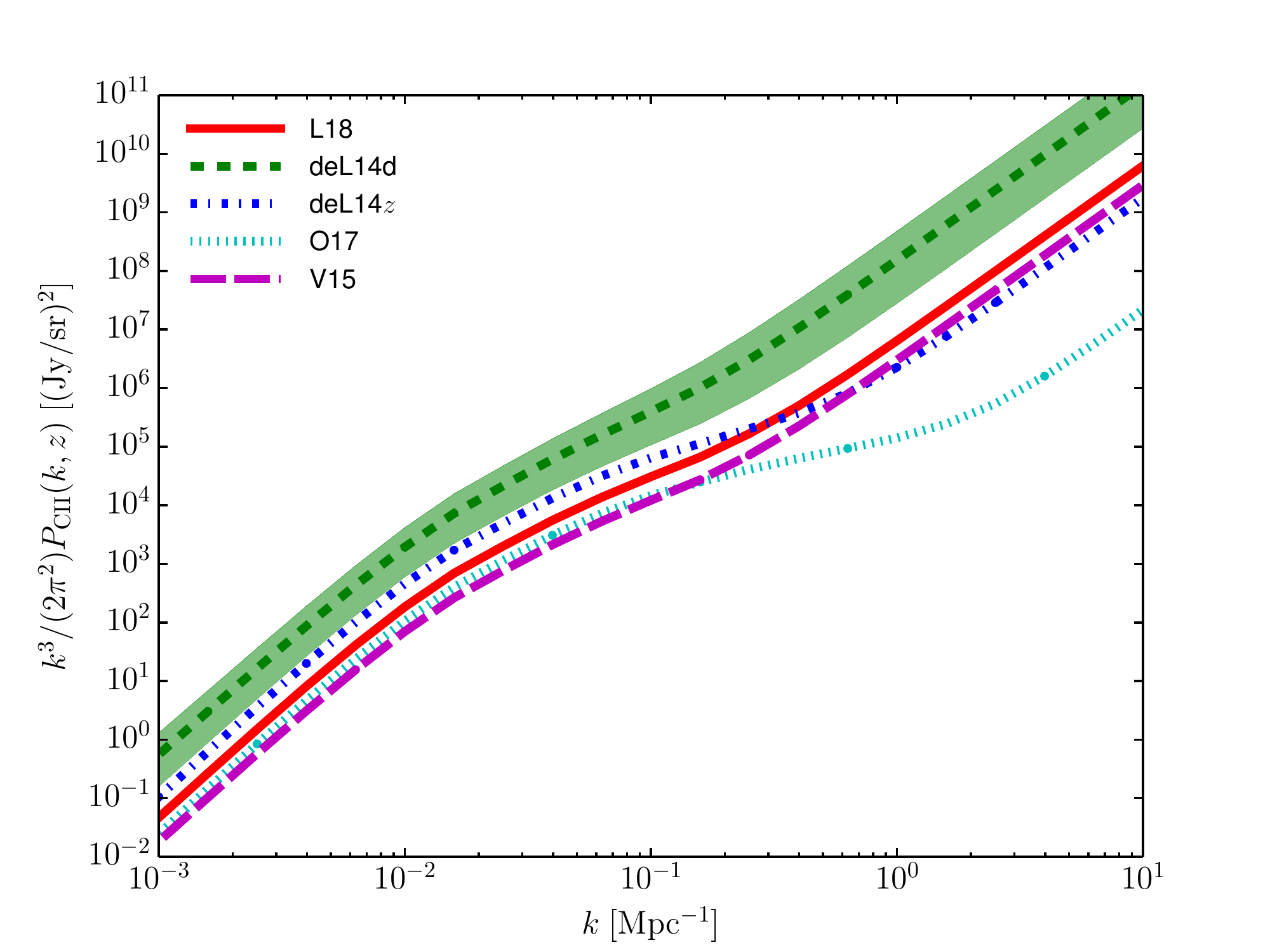}}
\subfigure{\includegraphics[width=90mm]{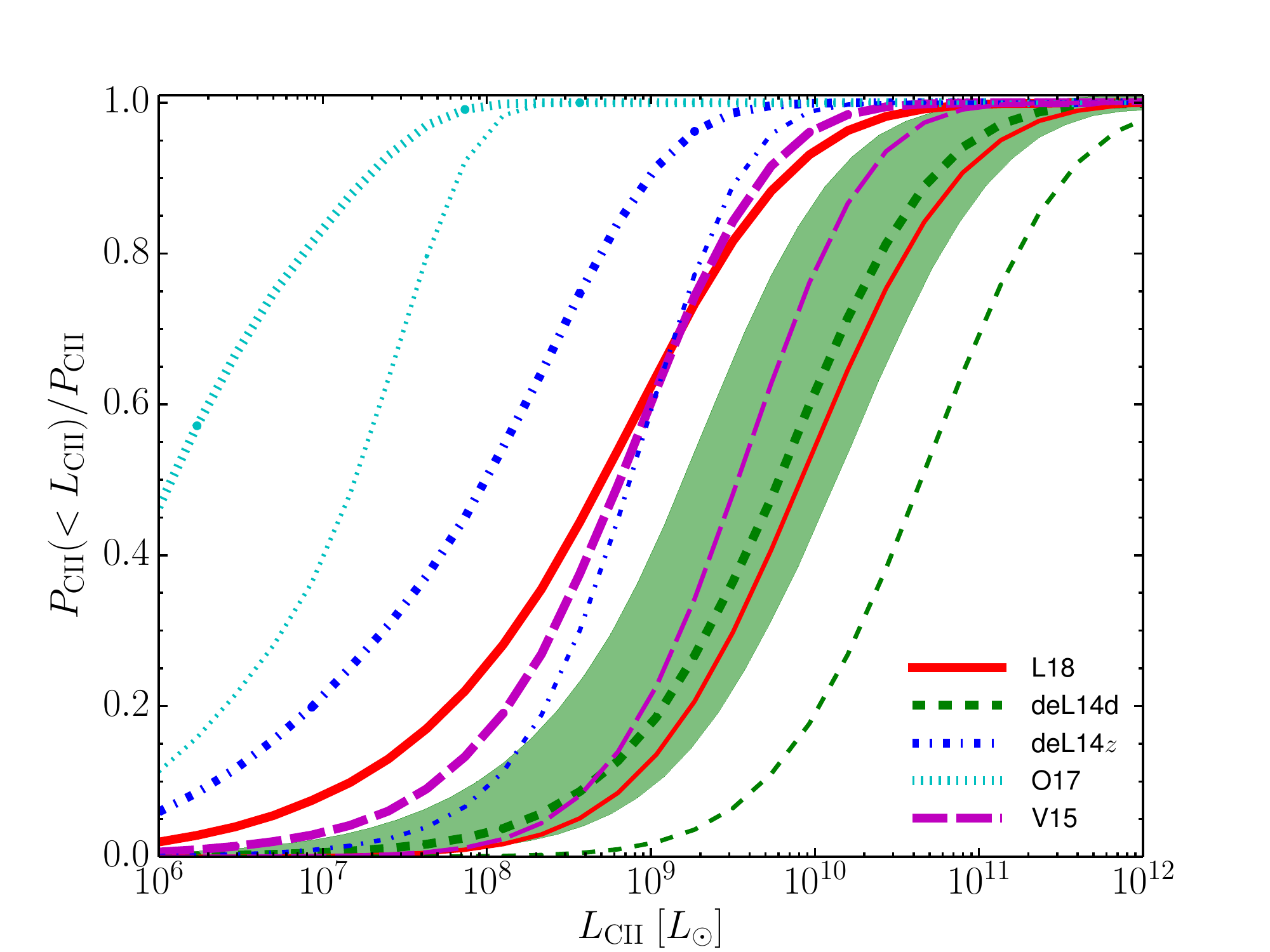}}
\caption{{\it Top:} \CII~power spectrum, including both clustering term and shot-noise contributions. {\it Bottom:} Fractional contribution from $<L_{\rm CII}$ galaxies to the \CII~power spectrum shot-noise (thin lines) and clustering (thick lines, taken at $k=0.1$ Mpc$^{-1}$) for the five considered $L_{\rm CII} - {\rm SFR}$ relations. For clarity we only show  uncertainties for the deL14d case. 
}
\label{fig:P_CII}
\end{figure}

In the top panel of Fig. \ref{fig:P_CII} we plot the \CII power spectrum, for the five $L_{\rm CII} - {\rm SFR}$ relations considered here. We find that the ``crossing scale" above which the shot-noise dominates over the clustering range from $\sim 0.2$ Mpc$^{-1}$ to $\sim2$ Mpc$^{-1}$. 
As this scale depends on the $L_{\rm CII} - {\rm SFR}$ relation, so it is not priori clear that $\sim k_{\rm max}$ is sampling the shot-noise. For simplicity, however, in the following we will refer to the power spectrum measured at $\sim k_{\rm max}$ as ``shot-noise". 

The fractional contribution from galaxies with luminosity smaller than $L_{\rm CII}$ to the shot-noise and clustering (measured at $k=0.1$ Mpc$^{-1}$) for the five considered $L_{\rm CII} - {\rm SFR}$ relations is shown in the bottom panel of Fig. \ref{fig:P_CII}. The key result is that, for these five $L_{\rm CII} - {\rm SFR}$ relations, with the only exception of O17, galaxies with $L_{\rm CII}\lsim10^{8-9}~L_\odot$ make a negligible contribution to the shot-noise. The O17 relation predicts instead that galaxies with $\lsim10^6~L_\odot$ contribute about $\sim10\%$ to the shot-noise signal, as a result of its shallower slope ($\gamma = 0.58$). Note that, however, the shot-noise absolute amplitude for O17 is the weakest among the five $L_{\rm CII}-{\rm SFR}$ relations. Compared with shot-noise, the fractional contribution of the clustering signal is larger at any $L_{\rm CII}$, i.e. the clustering signal is more sensitive to the faintest systems. deL14$z$ and O17 (L18 and V15)  predict that $\sim 50\%$ of the clustering signal comes from galaxies with $L_{\rm CII}\lsim10^8~L_\odot$ ($\lsim 10^9~L_\odot$), while for deL14d only $\sim20\%$ of clustering signal is from galaxies with  
$L_{\rm CII}\lsim10^9~L_\odot$.

\section{Results}

\subsection{Detectability of \CII power spectrum} 

\subsubsection{Signal-to-noise ratio} 

Guided by the telescope parameters given in Sec. \ref{telescopes}, we first assume a telescope with diameter $D= 6 $ m, $T_{\rm sys} = 50$ K, $N_{\rm det}=4000$ detectors and $\delta \nu_0=1$ GHz,
to survey a sky region $\Omega_{\rm survey}=(2~{\rm deg})^2$. 
For such an instrument,    $k_{\rm max}\approx1.7~{\rm Mpc^{-1}}$. 
We will only focus on the data from 237.6 to 271.5 GHz (corresponding to the \CII signal from $z = 6 -7$)
and assume the time dedicated to this frequency range is $t_{\rm obs}=1000$ hours. 
These choices are rather conservative, as usually surveys aim at larger areas/redshift ranges. For example, in its default configuration, the CCAT-p will survey the frequency range 190 - 450 GHz \citep{Vavagiakis2018}.  
We define this telescope/survey combination as the fiducial one. Its parameters are listed in Tab. \ref{tab:strategy} as ``S1".

\begin{figure}
\subfigure{\includegraphics[width=90mm]{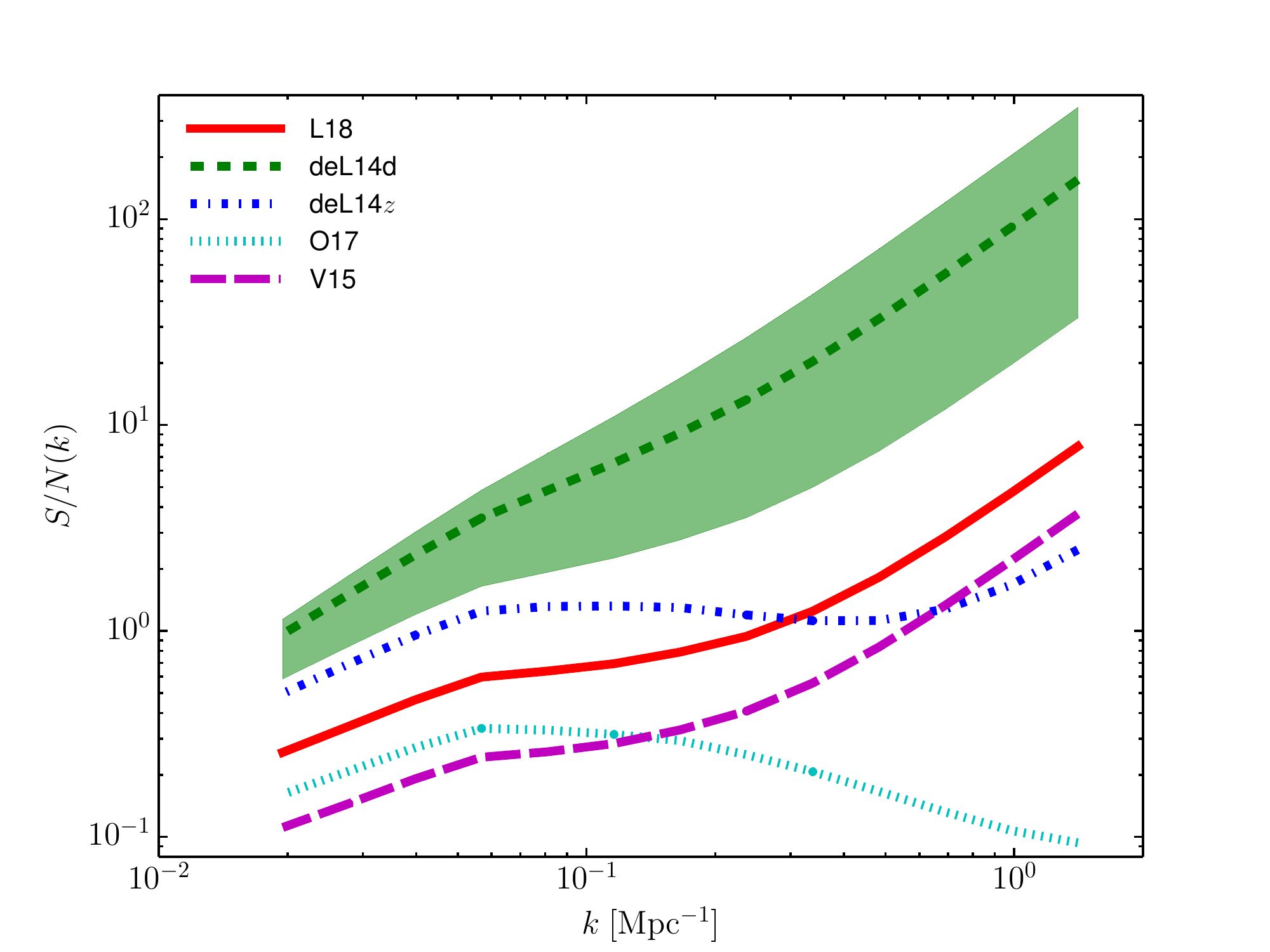}}
\subfigure{\includegraphics[width=90mm]{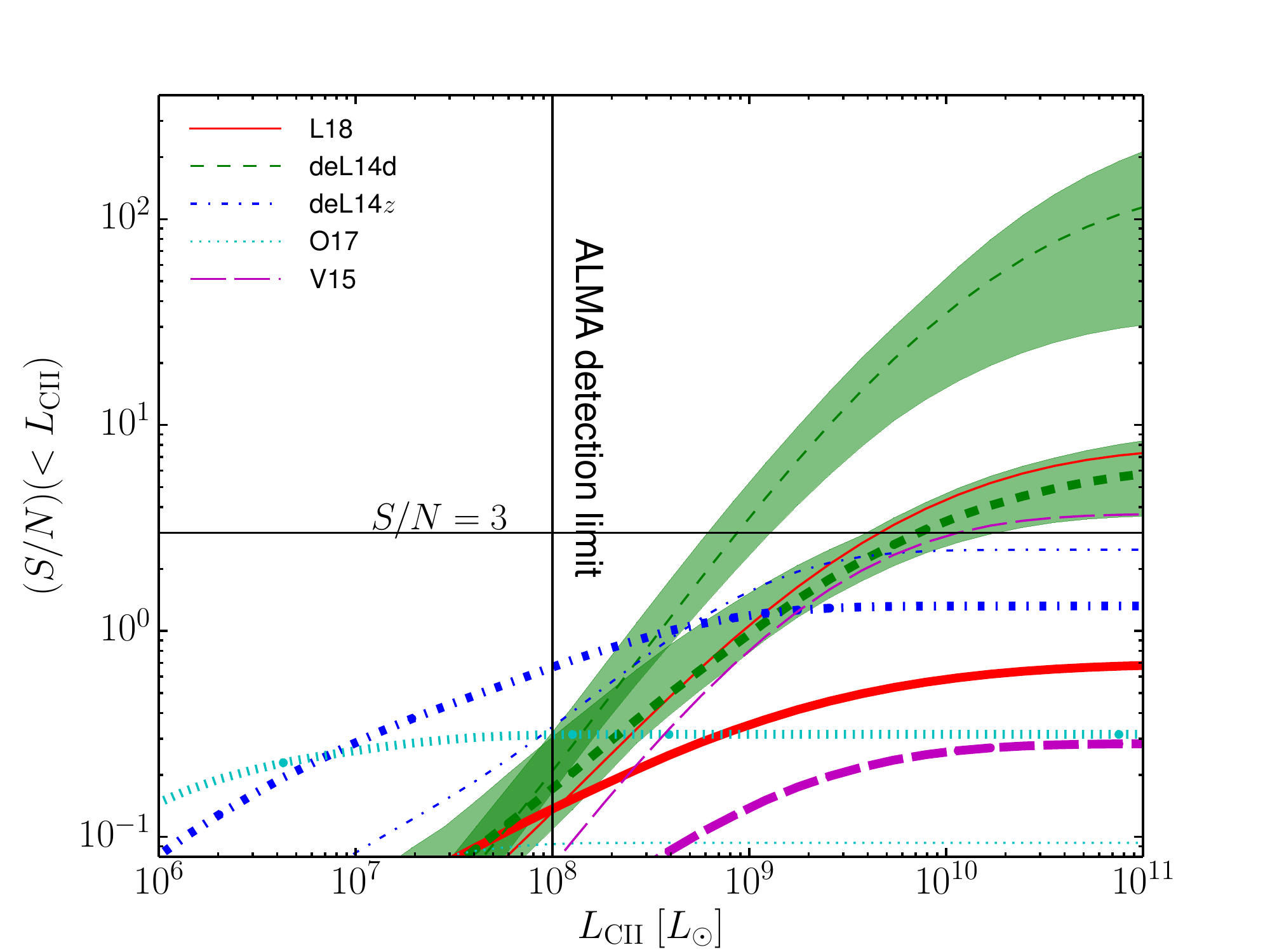}}
\caption{{\it Top:} The  signal-to-noise ratio as a function of $k$. {\it Bottom:} The $S/N$ of shot-noise (thin lines) and clustering (thick lines, at $\sim0.1$ Mpc$^{-1}$) of power spectrum from galaxies $<L_{\rm CII}$. To guide the eye, we have drawn approximate ALMA detection limit $\sim10^8~L_\odot$ and the $S/N=3$ line.
For clarity, uncertainties are plotted only for the deL14d case.
}
\label{fig:SNR}
\end{figure}

The $S/N$ (Eq. \ref{eq:SNR}) of the \CII power spectrum\footnote{We divide the $k=1.6\times10^{-2}-1.7~$Mpc$^{-1}$ range into 13 bins of width $d{\rm ln}k =0.3$.} from all luminosities is shown in the top panel of Fig. \ref{fig:SNR}.  
In general the shot-noise has a larger $S/N$ with respect to the clustering term (with O17 being an exception). 

In the bottom panel of Fig. \ref{fig:SNR} we plot the $S/N$ of the shot-noise (thin lines) and clustering (thick lines, at $ k=0.1$ Mpc$^{-1}$) of \CII power spectrum from galaxies with luminosity $< L_{\rm CII}$.  
Both the shot-noise and clustering signals are undetectable  based on the deL14$z$ and O17 $L_{\rm CII}-{\rm SFR}$ relations ($S/N<3$). For deL14d, however, both the shot-noise ($\gsim9\times10^8~L_\odot$) and clustering signals ($\gsim8\times10^9~L_\odot$) are detectable, while for L18 only the shot-noise from galaxies $\gsim5\times10^9~L_\odot$ is detectable.
For V15, only the shot-noise from galaxies with $L_{\rm CII}\gsim10^{10}~L_\odot$ is marginally detectable. 
In principle, then, IM experiments might bring important information on the $L_{\rm CII}-{\rm SFR}$ relation for high-$z$ galaxies. 
 
Note that, given that the ALMA detection limit is $\sim 10^8 L_\odot$, 
for the five $L_{\rm CII}-{\rm SFR}$ relations considered here, galaxies that are too faint to be detected by ALMA can neither be detected with the fiducial IM experiment assumed here. In spite of this, an IM mapping experiment could provide substantial information on the bright-end of the LF sampling a population of massive, rare galaxies which would be basically impossible to find given the small field of view of ALMA observations. 

For example, among the five $L_{\rm CII}-{\rm SFR}$ relations considered, the deL14d model has the largest fractional contribution from bright galaxies. In this model, about half of the clustering signal is from galaxies brighter than $\sim5\times10^9~L_\odot$ 
(for shot-noise, half of the signal is from galaxies brighter than $\sim5\times10^{10}~L_\odot$).
By looking at the \CII LF in Fig. \ref{fig:CIILF}, we see that galaxies with $L_{\rm CII}\gtrsim 5\times10^9~L_\odot$ have a number density of $\sim10^{-5}$ Mpc$^{-3}$. To detect, say, $\sim$10 such galaxies, the survey volume should be $\gtrsim10^6$ Mpc$^3$, much larger than the typical volumes surveyed by ALMA. 

Such prediction is based on the assumption that the adopted $L_{\rm CII}-{\rm SFR}$ relation holds in the entire SFR range of interest.
It is not clear whether this is true. However, at low redshift, there are indeed star-forming galaxies with \CII luminosities up to $\sim10^{10}~L_\odot$ detected (e.g. \citealt{Stacey2010,Wagg2012,Magdis2014,Brisbin2015}). 

It is also possible that, due to feedback effects, bright galaxies are less efficient  \CII emitters \citep{Rybak2019, Pallottini2019}. In this case, the number of \CII bright galaxies would be smaller than we predict based on a single $L_{\rm CII}-{\rm SFR}$.  \CII IM observations could clarify this issue.  

On the other hand, galaxies $\gtrsim5\times10^9~L_\odot$ should have an UV magnitude $\lesssim-22$. Such galaxies are detectable in  UV surveys. For example,  \citet{Bowler2015} have detected $\lesssim 20$ galaxies brighter than $-22$ with $5.7 <z<6.3$ in a 1.65 deg$^2$ sky area. However, puzzingly, \citet{Willott2013} found a smaller number of bright galaxies in a larger area (4 deg$^2$). The cross-correlation between the \CII IM and UV-detected galaxies would be useful for providing solid information of \CII radiative processes in bright galaxies.

\begin{figure}
\subfigure{\includegraphics[width=90mm]{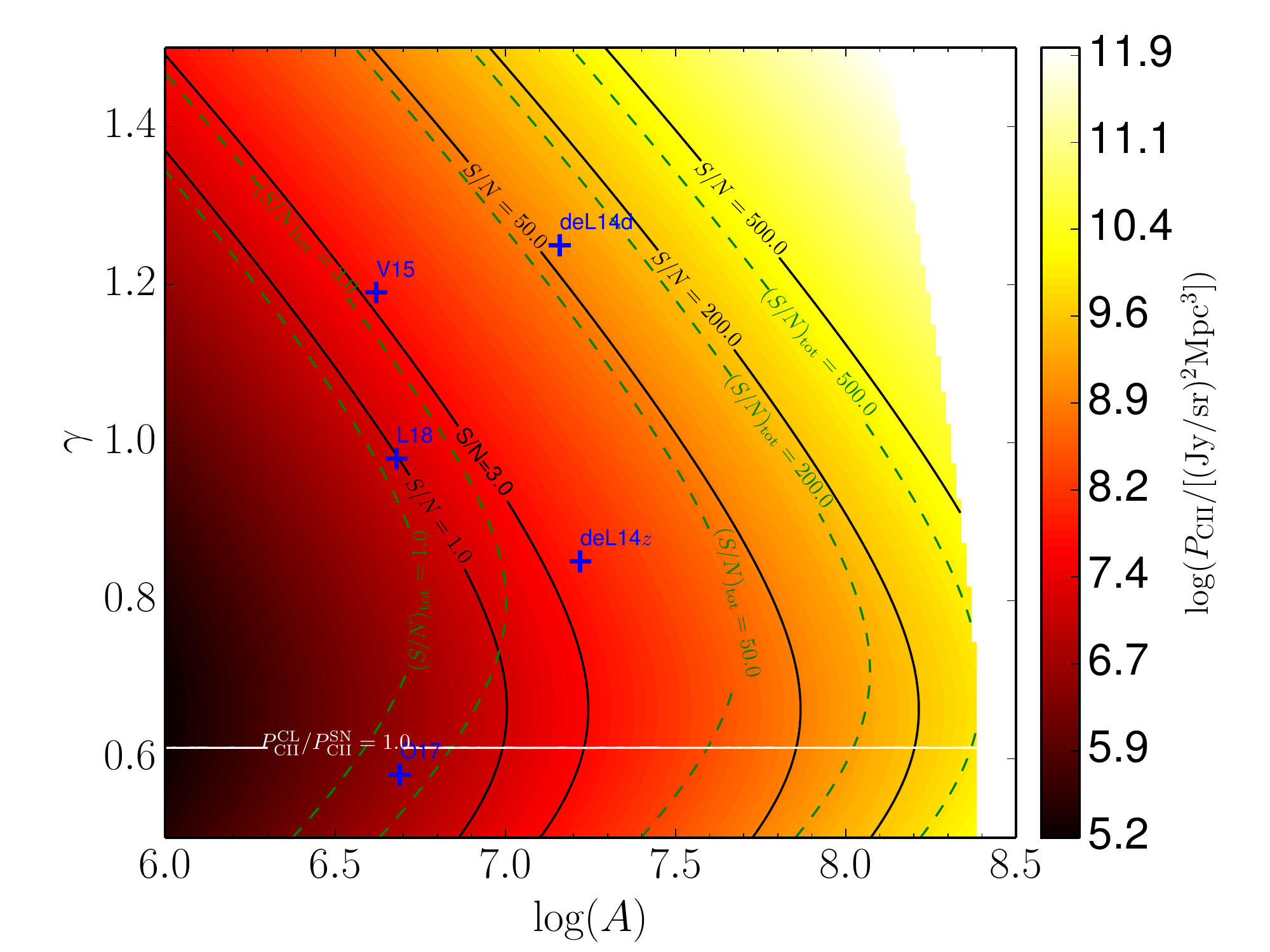}}
\caption{
The \CII power spectrum at $k_{\rm max}$ (colored contours), the $S/N$  (solid line contours) at $k_{\rm max}$, and the total $(S/N)_{\rm tot}$ (dashed contours) as a function of ${\rm log}A$ and $\gamma$.  We have excluded the parameter space that would produce a \CII~LF larger than any of the three upper limits in \citet{Yamaguchi2017}. We mark the regions with $P_{\rm CL}/P_{\rm SN}=1.0$ by a horizontal line ($\gamma \sim 0.6$); below this line the shot-noise may be contaminated by the clustering signal at $k_{\rm max}$.  We also mark the positions of the L18, deL14d, deL14$z$, O17  and V15 $L_{\rm CII}-{\rm SFR}$ relations.
}
\label{fig:logP_CII}
\end{figure}

Given the present uncertainties on the high-$z$ \CII LF, it makes sense to explore a wider region of the $\log A -\gamma$ parameter space for the fiducial S1 survey, $6<{\rm log}A<9$ and $0.5<\gamma<1.5$, than the five $L_{\rm CII}-{\rm SFR}$ relations. The outcome of this generalized study is summarised\footnote{In the 2D contour map we ignore uncertainties in the observed UV LF, $\beta$-$M_{\rm UV}$ relation and IRX-$\beta$ relation, and adopt $\sigma_L=0.4$ dex.} in Fig. \ref{fig:logP_CII} in terms of the resulting shot-noise, $S/N$ evaluated at $\sim k_{\rm max}$,  and $(S/N)_{\rm tot}$. 

The shot-noise signal is detectable in the region where $S/N>3$, or $P_{\rm CII}(k_{\rm max})\gsim10^{7.7}~{\rm (Jy~sr^{-1})^2Mpc^{3}}$. Clearly, large amplitudes, log$A$, and steep slopes, $\gamma$, of the $L_{\rm CII}-{\rm SFR}$ relation make the detection easier. Note that for $\gamma \lsim 0.6$ the clustering signal dominates over the shot-noise at least at $k\lsim k_{\rm max}=1.7$ Mpc$^{-1}$. In this case, a detection is possible even from galaxies for which $P_{\rm CII}(\sim k_{\rm max})\gsim10^{6.7}~{\rm (Jy~sr^{-1})^2Mpc^{3}}$,
because of the clustering contribution to the total signal. 

\subsubsection{Detection depth and number counts.}\label{sec:num_counts}

Although the IM signal contains the record of the collective emission from all galaxies, including the faintest ones, the previous results show that for most $L_{\rm CII}-{\rm SFR}$ relations the dominant  contribution to the IM signal (shot-noise or clustering) comes from galaxies that are already well at reach of current optical/IR surveys (and the more so with the advent of JWST).
 
\begin{figure}
\subfigure{\includegraphics[width=90mm]{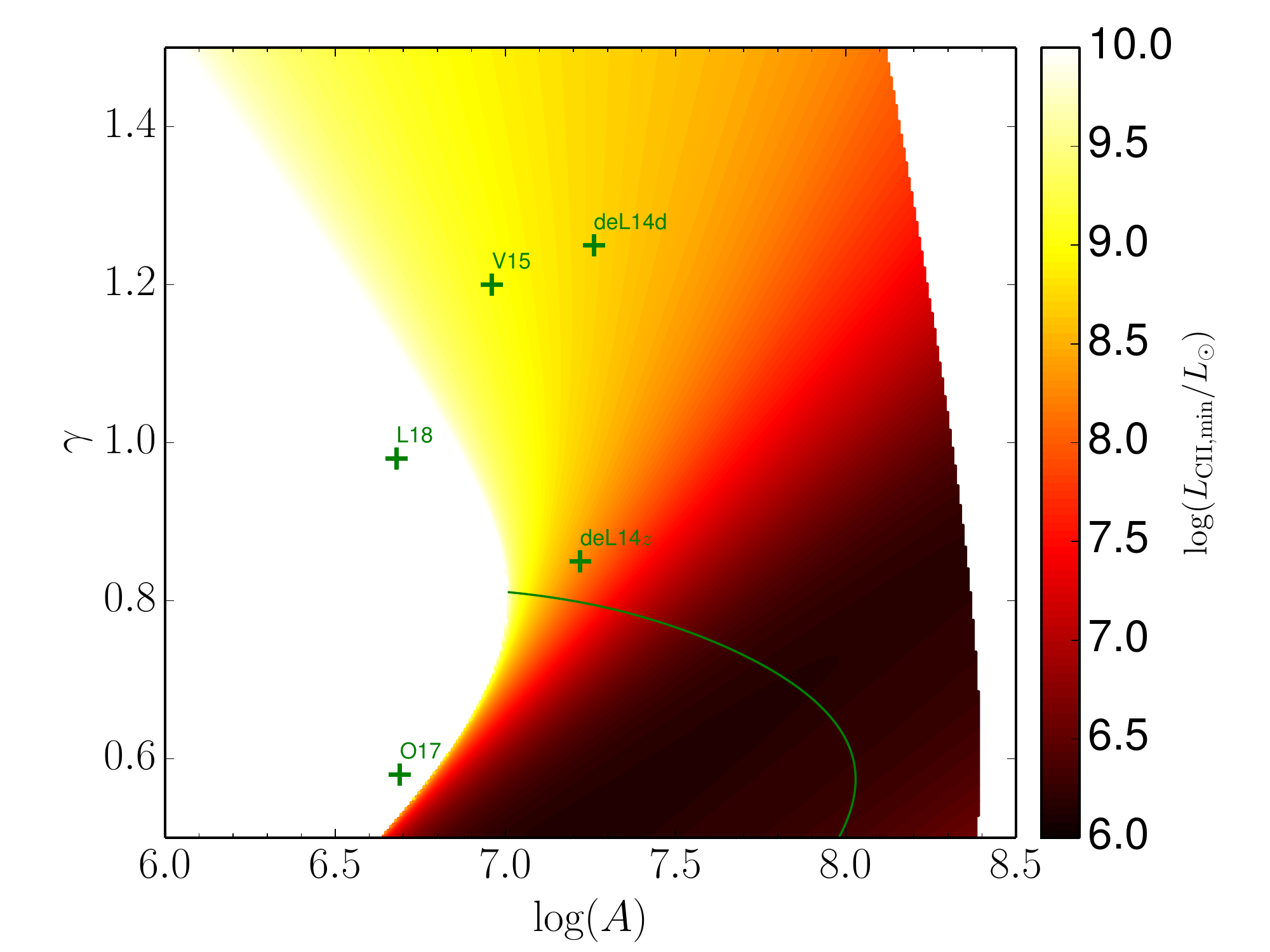}}
\subfigure{\includegraphics[width=87mm]{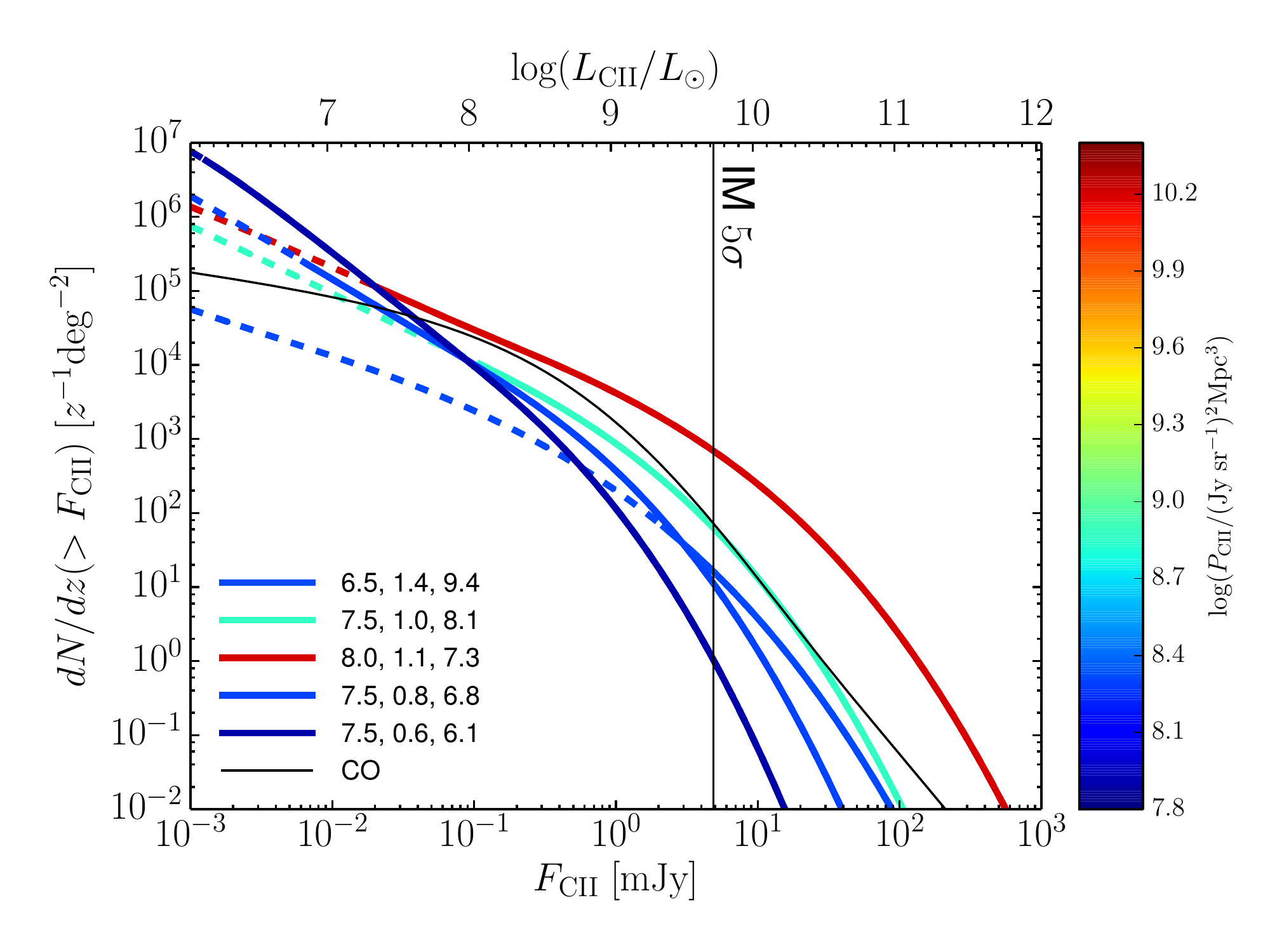}}
\caption{
{\it Top:} \CII luminosity of the faintest sources, $L_{\rm CII,min}$, detectable by our fiducial IM experiment as a function of the $\log A$ and $\gamma$ parameters of the $L_{\rm CII} - {\rm SFR}$ relation, defined as Eq. (\ref{eq:LCIImin}). In regions below (above) the solid green curve the signal is mainly contributed by clustering (shot-noise). We have excluded regions where $(S/N)_{\rm tot} < 3$ (bottom left), and regions that break the \citet{Yamaguchi2017} upper limit constraints (top right).
{\it Bottom:} \CII number counts for different values of ($\log A$, $\gamma$, log$L_{\rm CII,min}$) as given in the legend. The color of each curve gives the corresponding value of $P_{\rm CII}(k_{\rm max})$. Number counts below (above) $L_{\rm CII,min}$ are plotted as dashed (solid) lines. The vertical line indicates $5\sigma$ of the IM noise, this criterion is generally used for resolving point sources from a map. As a comparison, the thin solid curve refers to the number counts from CO lines (for all detectable rotational transitions, see \citealt{Yue2015}). 
}
\label{fig:logLCIImin}
\end{figure}

Ideally, one can claim detection of the faint galaxies contribution as long as their total signal-to-noise ratio is $> 3$.
Consider a $L_{\rm CII,min}$ for which
\begin{align}
(S/N)_{\rm tot}(<L_{\rm CII,min},k,z)&=3.
\label{eq:LCIImin}
\end{align}
We define $L_{\rm CII,min}$  as the detection depth of the considered IM experiment; this quantity is shown in the top panel of Fig. \ref{fig:logLCIImin}. Again we adopt $\sigma_L=0.4$ dex.
We have excluded regions where $(S/N)_{\rm tot} < 3$  for all \CII luminosities (the left empty patch), and regions that break the \citet{Yamaguchi2017} upper limit constraints. 
In regions below (above) the solid curve in the panel the signal is mainly contributed by clustering (shot-noise). 

We find that $L_{\rm CII,min}$ is almost always smaller than the $5\sigma$ of the IM noise, $\sim10^{9.7}~L_\odot$. 
Since 5$\sigma$ criterion is generally used for point source identification, for an IM experiment its power spectrum detection depth is usually deeper than its point sources detection depth. 
Surprisingly, for some (large) log$A$ and (small) $\gamma$, $L_{\rm CII,min}$ could be as small as $\lsim10^7~L_\odot$. This is because when $\log A$ is large, the galaxy is brighter in \CII. However, when $\gamma$ is smaller then the \CII~LF drops very fast at the bright-end; as a result the majority of the signal is from the faint galaxies, see the O17 curve in Fig. \ref{fig:CIILF}.  Therefore, whether an IM experiment is useful for detecting {ALMA}-unresolved galaxies depends on the slope of the $L_{\rm CII}-{\rm SFR}$ relation. 

As a final step, we compute the cumulative  number counts per unit area per unit redshift (i.e. the surface number density) of galaxies above a given flux, $dN/dz(> F_{\rm CII})$. To this aim we need first to transform the \CII~luminosity into a \CII flux by rewriting Eq. (\ref{eq_flux}) as follows,
\begin{equation}
F_{\rm CII}=\frac{L_{\rm CII}}{4\pi d_L^2} \frac{1}{\Delta \nu_0},
\label{eq_flux2}
\end{equation}
further assuming that the spectral resolution of the telescope, $\Delta \nu_0$, is much larger than the line width.  
It follows that 
\begin{align}
\frac{dN}{dz}(>F_{\rm CII})&= r^2\frac{dr}{dz}   \int_{{\rm log}L_{\rm CII}(F_{\rm CII})}^\infty \frac{dn}{d{\rm log}L_{\rm CII}} d{\rm log}L_{\rm CII}. 
\label{eq:dNdz}
\end{align}
 
To help intuition and illustrate how the shape of \CII LF is related to different detection limits, in the bottom panel of Fig. \ref{fig:logLCIImin} we plot the \CII number counts for some selected $\log A$ and $\gamma$ values. Smaller $\gamma$ produces a steeper number count faint-end slope, while larger 
$\log A$, as expected, increases the number counts.
 
\subsubsection{Recovering the \CII LF}
\begin{figure}
\subfigure{\includegraphics[width=90mm]{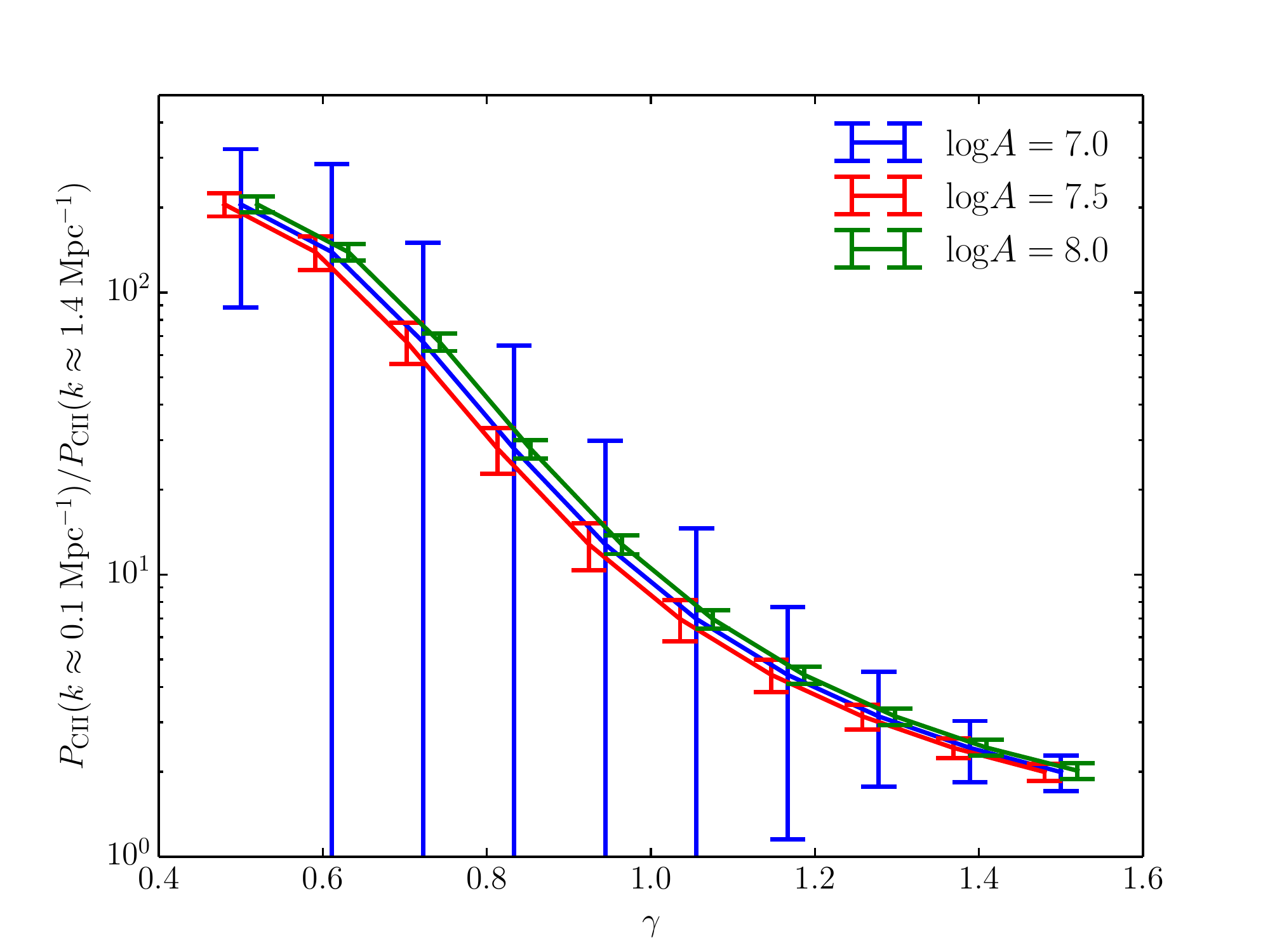}}
\caption{The $P_{\rm CII}(k\approx0.1~{\rm Mpc}^{-1})/P_{\rm CII}(k\approx1.4~{\rm Mpc}^{-1})$ ratio (i.e. a proxy for the clustering/shot-noise relative contribution) vs. $\gamma$, for log$A=7.0$, 7.5 and 8.0, respectively. 
}
\label{fig:PCL2PSN}
\end{figure}

The detection of the \CII power spectrum signal will allow to reconstruct the \CII LF.
However, as we have seen above, measuring the shot-noise only does not provide separate constraints on the two parameters determining the LF, log$A$ and $\gamma$ (see, e.g. Fig. \ref{fig:logP_CII}). Luckily, as the clustering and shot-noise terms have a different dependence on these two parameters, the full power spectrum can break such degeneracy.
In  Fig. \ref{fig:PCL2PSN} we show the $P_{\rm CII}(k\approx0.1~{\rm Mpc}^{-1})/P_{\rm CII}(k\approx1.4~{\rm Mpc}^{-1})$ ratio as a function of $\gamma$. Such ratio is a proxy for the clustering/shot-noise relative contribution. As long as the $S/N$ is sufficiently large, this ratio can be used to precisely measure $\gamma$, as it is rather insensitive to the precise value of log$A$, see Fig. \ref{fig:PCL2PSN}. As a caveat, though, recall that  errors become larger as $A$ decreases.

\begin{figure*}
\subfigure{\includegraphics[width=80mm]{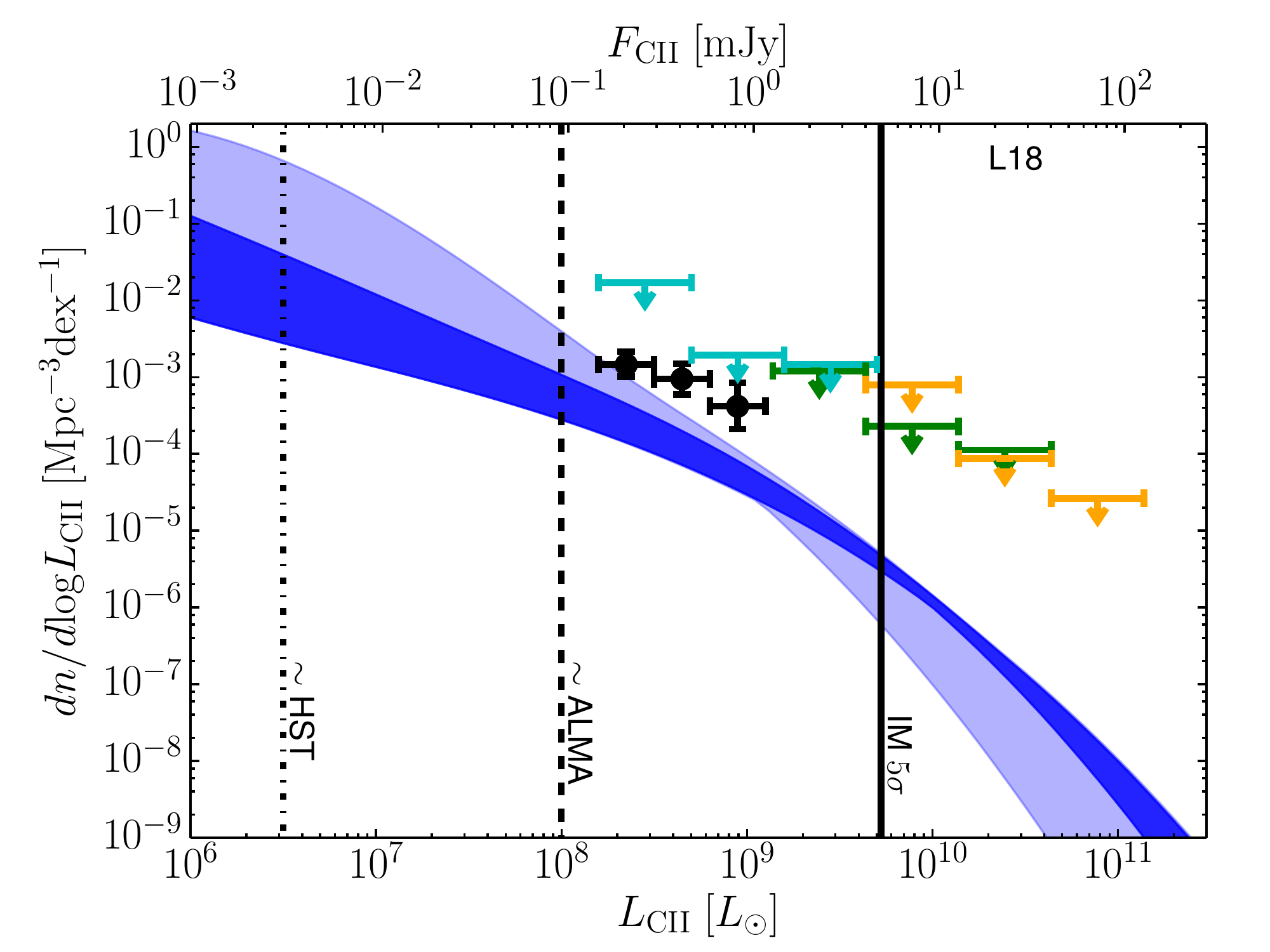}}
\subfigure{\includegraphics[width=80mm]{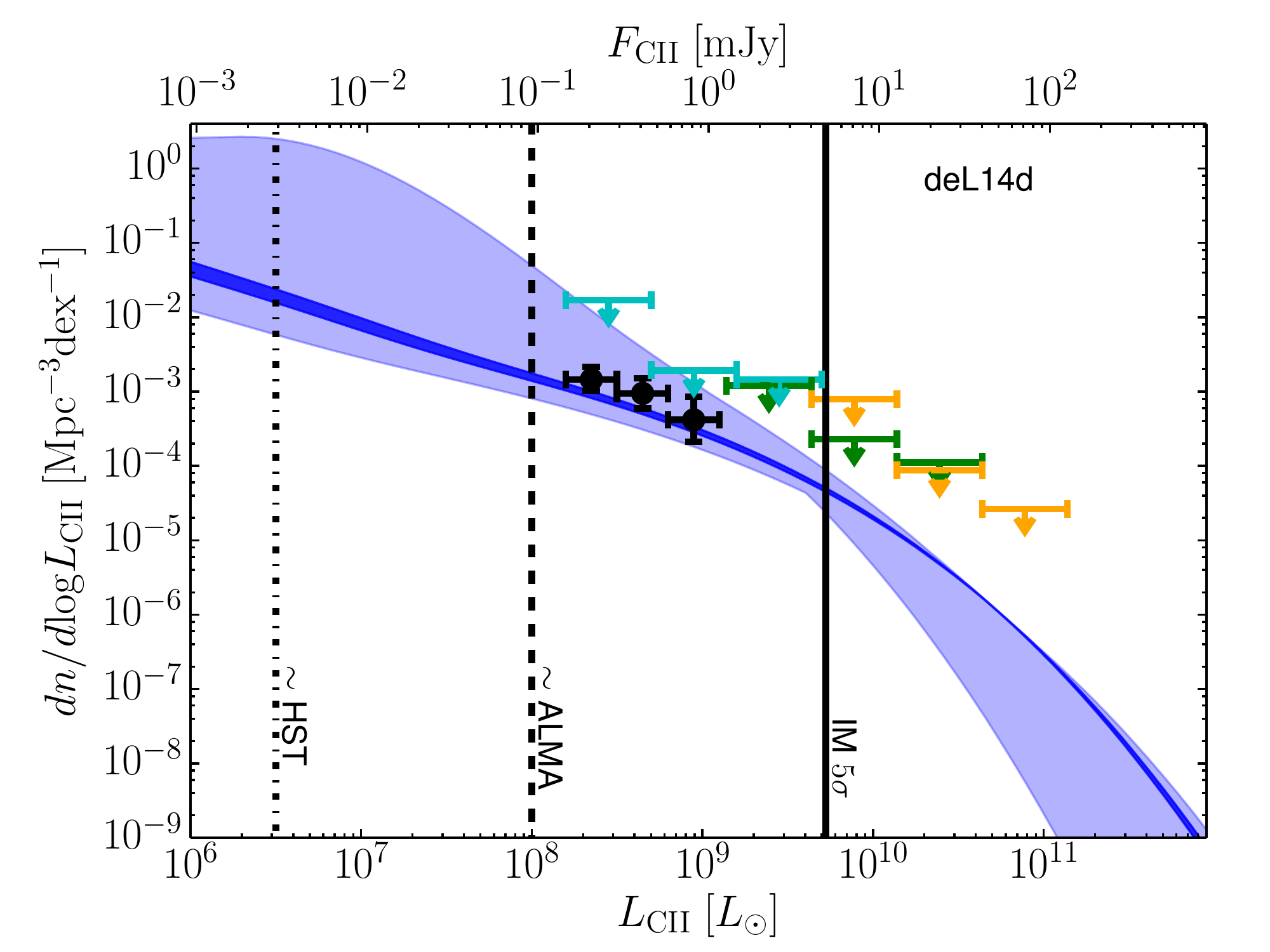}}
\subfigure{\includegraphics[width=80mm]{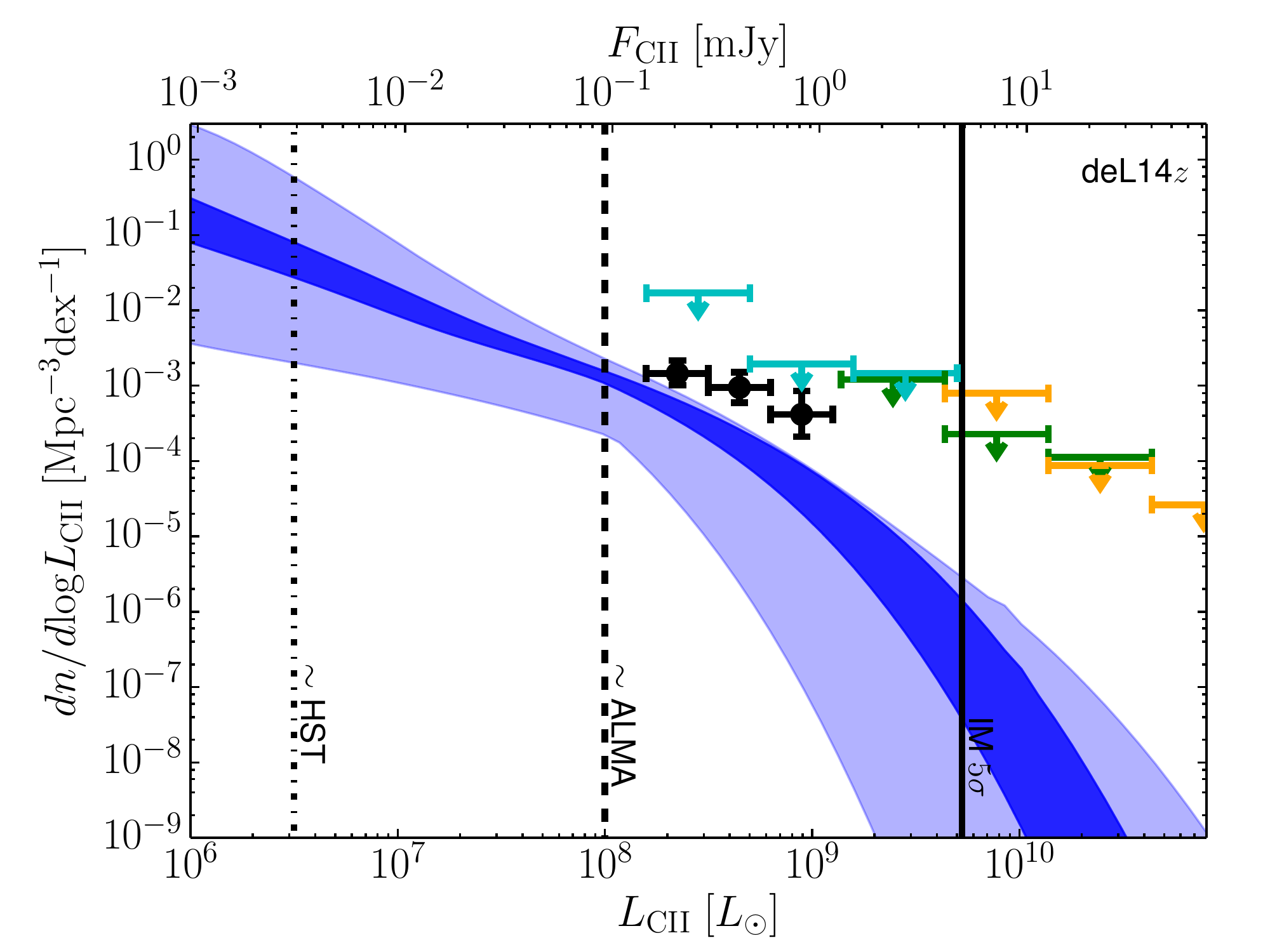}}
\subfigure{\includegraphics[width=80mm]{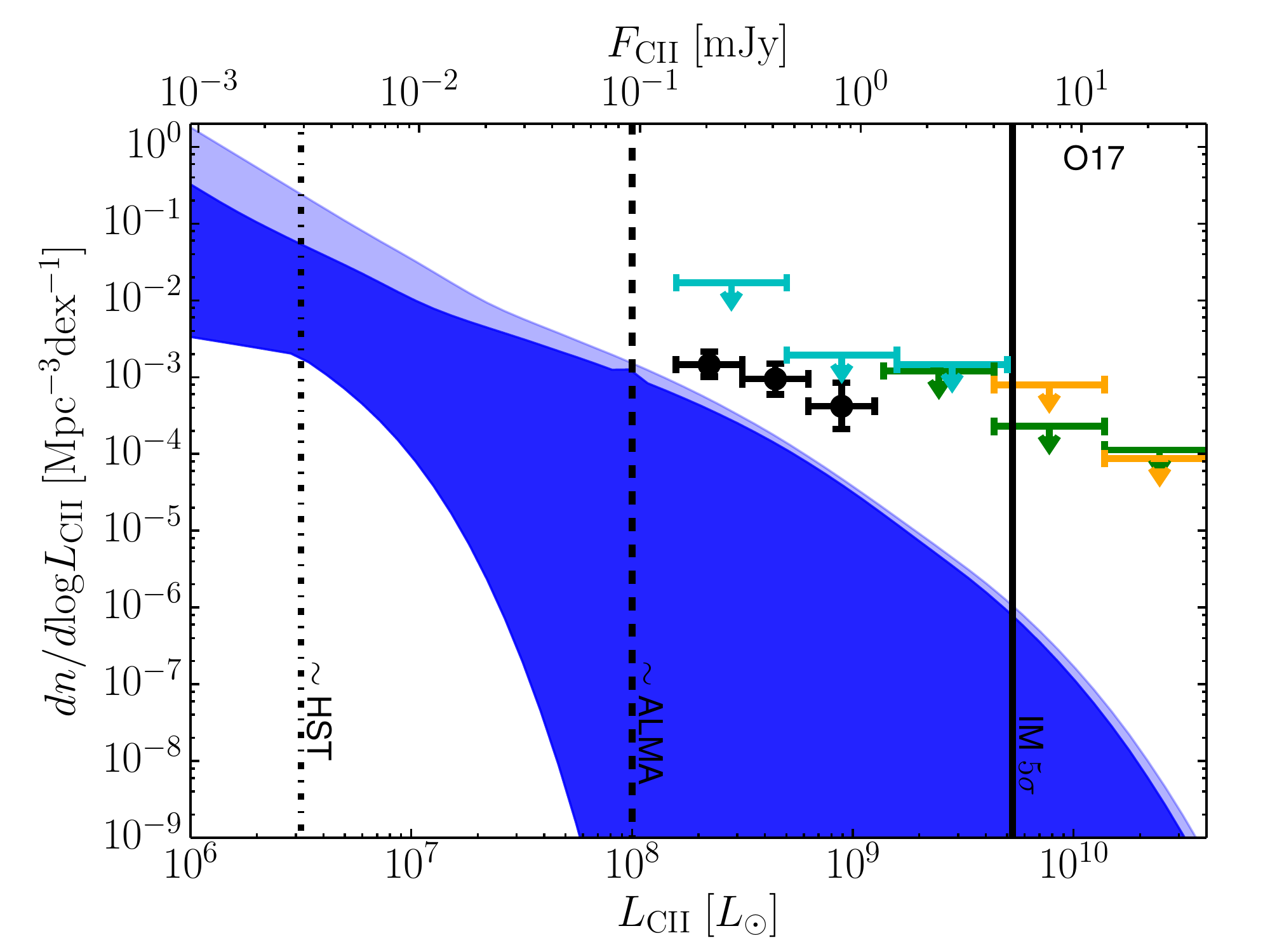}}
\subfigure{\includegraphics[width=80mm]{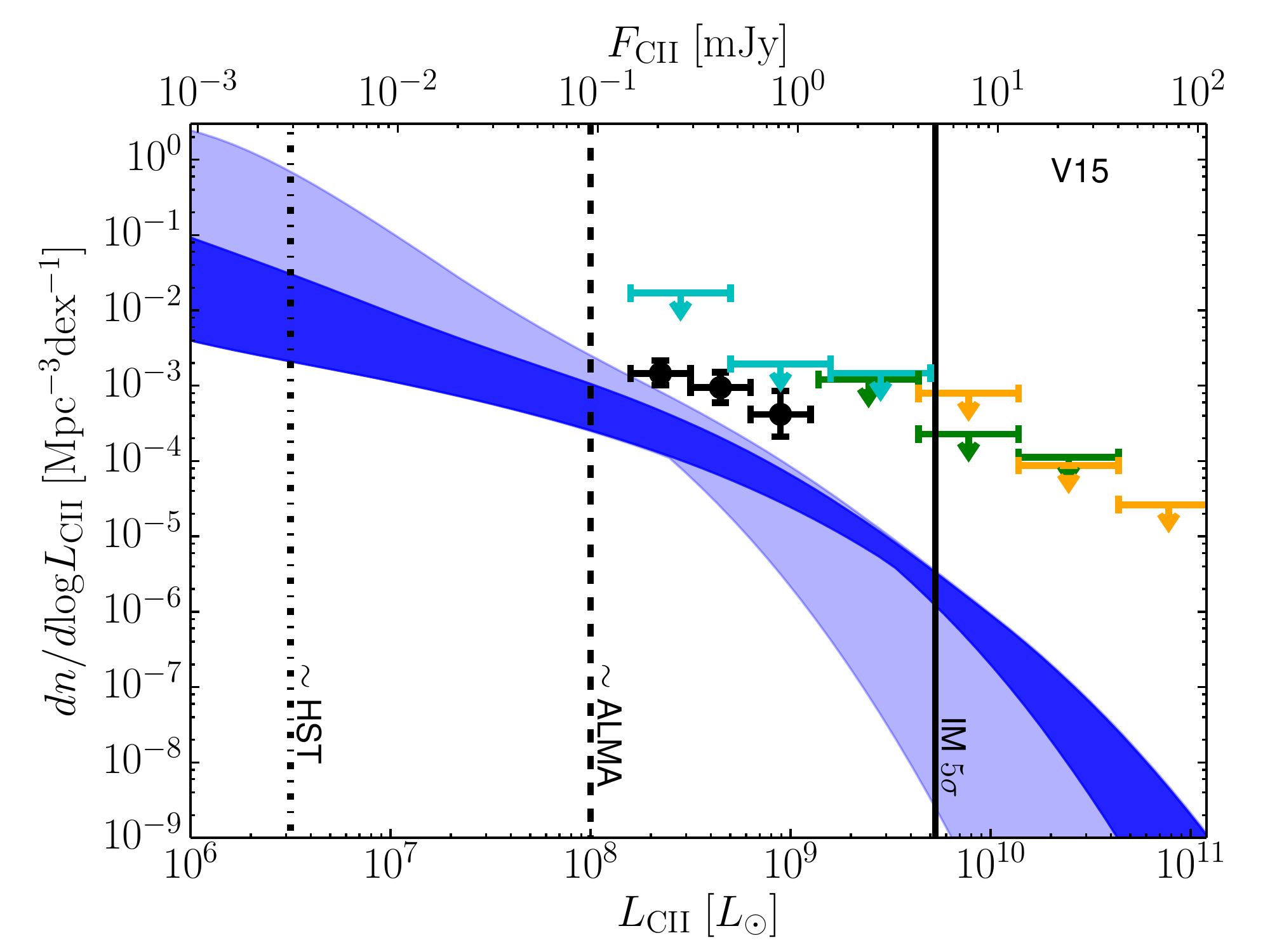}}
\caption{Recovered \CII~LF envelopes corresponding to the $1\sigma$ uncertainties of ${\rm log}A$ and $\gamma$ at $z\sim6$ from the \CII power spectrum in the fiducial survey. The L18, deL14d, deL14$z$, O17 and V15 $L_{\rm CII}-{\rm SFR}$ relations are assumed for generating the input power spectrum. In each panel the regions with dark (light) color refer to using the full power spectrum (shot-noise only). We use $\delta \nu_0$=1.0 GHz to translate the \CII~luminosity into the flux per beam, see the top $x$-axis. We also mark with vertical lines the detection limits by {\it HST} ($M_{\rm UV}\sim-16.5$, dashed-dotted), {ALMA} ($\sim10^8~L_\odot$,dashed) and the $5\sigma$ level of the underlying IM survey (solid), respectively.}
\label{fig:dn_dlogLCII_UVLF2}
\end{figure*}

\begin{figure}
\subfigure{\includegraphics[width=80mm]{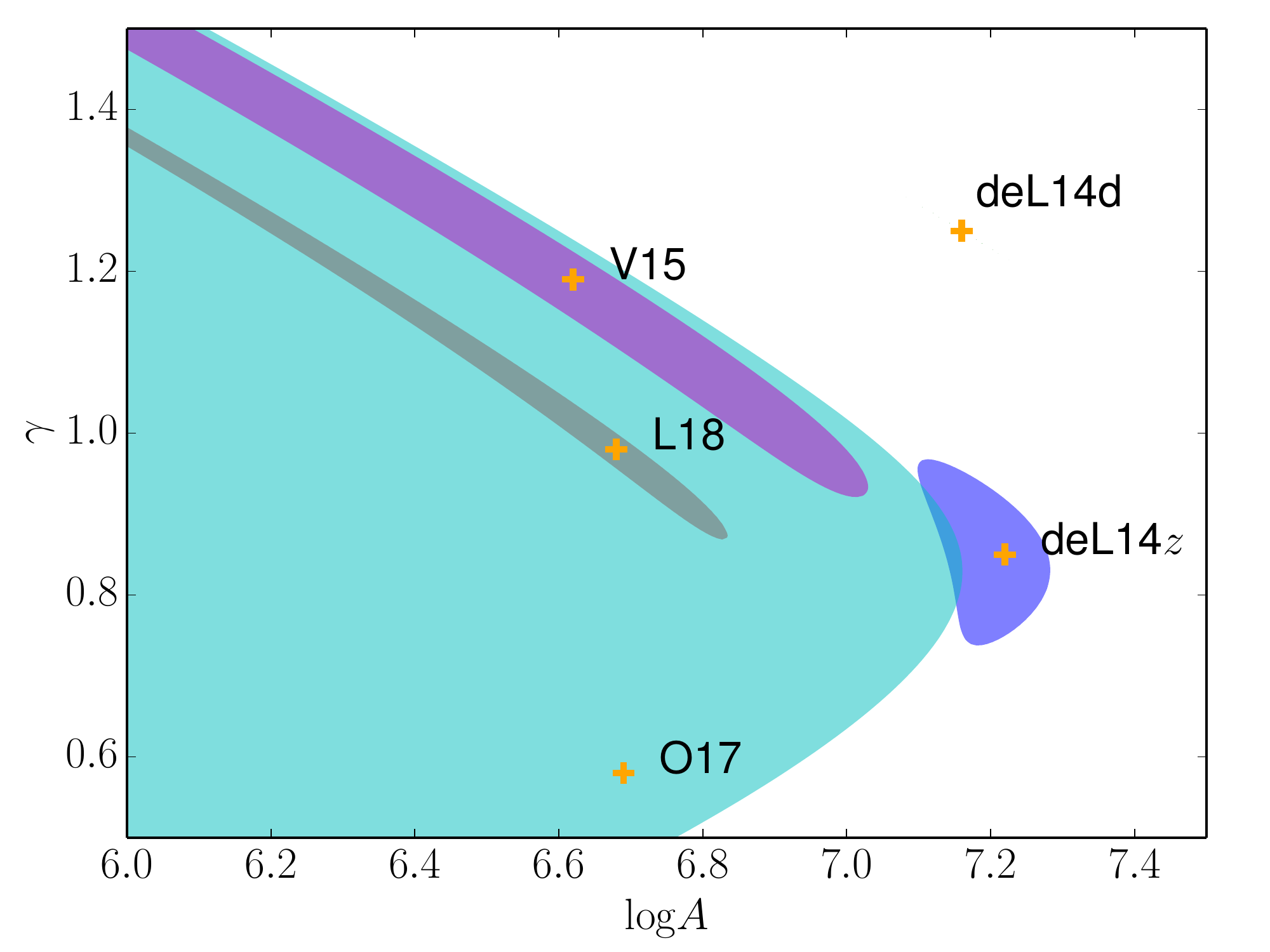}}
 \caption{The 1$\sigma$ constraint contours of the ${\rm log}A-\gamma$ parameters from the full power spectrum for the five  $L_{\rm CII}-{\rm SFR}$ relations. The ``+" symbols mark the input values.}
\label{fig:logA_gamma}
\end{figure}

In Fig. \ref{fig:dn_dlogLCII_UVLF2} we plot the \CII LF envelopes corresponding to 1$\sigma$ uncertainties on log$A$ and $\gamma$ constrained by the shot-noise only (light color) and by the full power spectrum (dark color), for the fiducial survey S1. For the O17 models, the \CII power spectra have low amplitudes and the instrumental noise dominates the error budget. In this case, using the full power spectrum does not significantly improve the LF accuracy. However, if the instrumental noise is sub-dominant, as in the L18, deL14d, deL14$z$ and V15 cases, then the full power spectrum provides an obvious advantage by shrinking the uncertainty envelopes.
Fig. \ref{fig:logA_gamma} shows their  1$\sigma$   ${\rm log}A- \gamma$  constraint contours\footnote{As the deL14d model contour is too small, for graphical reasons we only mark it with a cross.}.

Ideally, this is the key result that we can expect from our 
fiducial like IM experiment combined with the existing upper limits on the LF coming from pointed observations. 
We warn that these results depend on the underlying assumption of a single power-law form for $L_{\rm CII}$-SFR relation,  ${\rm log}L_{\rm CII}={\rm log}A+\gamma{\rm SFR}\pm\sigma_L$,  and shown in Fig. \ref{fig:logLCII-MUV}. 
At present this remains only an educated guess for high redshift galaxies, as also hinted by the more complex shape found by state-of-the-art simulations in P15, and more recently by \citealt{Pallottini2019}.
ALMA surveys covering a larger area and/or reaching to fainter fluxes, plus a more accurate determination of the galaxy SFR, would be of fundamental importance for this issue.

\subsection{Survey strategies}
 
We finally investigate how to optimize a survey for a \CII IM experiment. Generally, given the observation time and bandwidth, when $P_N \gtrsim P_{\rm CII}$, $S/N\propto 1/\sqrt{\Omega_{\rm survey}}$, a narrow and deep field has larger $S/N$. However, when $P_N \lesssim P_{\rm CII}$, $S/N\propto \sqrt{\Omega_{\rm survey}}$, wider field has larger $S/N$. For some applications, though, a highest $S/N$ is not necessarily the first priority indicator for optimizing surveys; instead, a wide survey is preferred in order to sample long wavelength $k$-modes carrying cosmological information and break parameters degeneracy, even at the price of lowering the $S/N$.

\begin{figure}
\subfigure{\includegraphics[width=90mm]{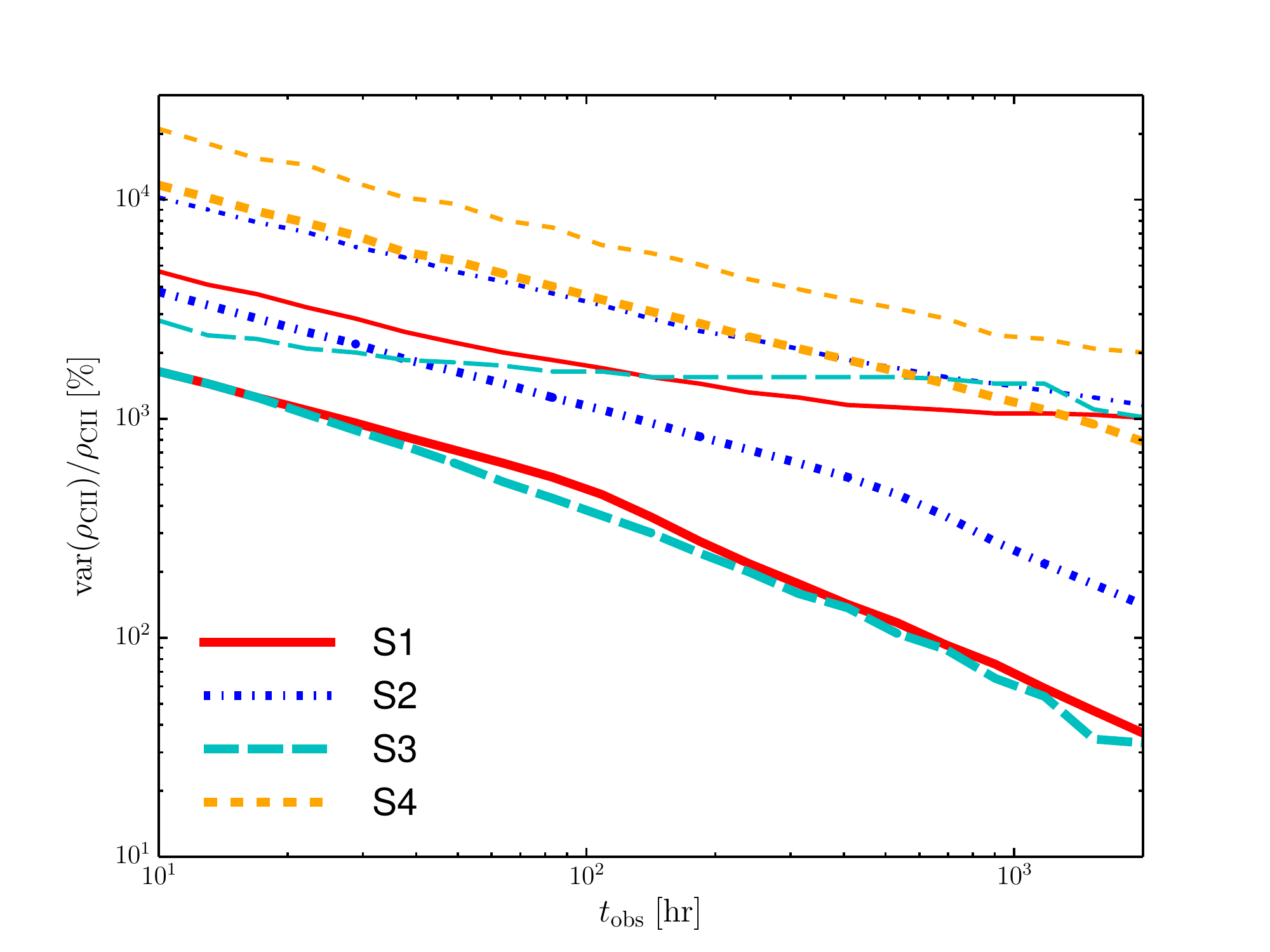}}
\caption{
The  var($\rho_{\rm CII}$) relative to $\rho_{\rm CII}$ as a function of $t_{\rm obs}$,  for the four surveys listed in Tab. \ref{tab:strategy}.  
Thin lines refer to the \CII luminosity density recovery using the shot-noise only, while the thick lines refer to use of the 
full \CII power spectrum. 
}
\label{fig:strategies}
\end{figure}

We discuss this issue in the context of recovering a fundamental physical quantity as the \CII luminosity density\footnote{ The flux is $I_{\rm CII}(z)=[c/4\pi \nu_{\rm CII}H(z)]\rho_{\rm CII}(z)$.
}, the first moment of the LF:
\begin{equation}
\rho_{\rm CII}(z)=\int L_{\rm CII} \frac{dn}{d{\rm log}L_{\rm CII}} d{\rm log}L_{\rm CII}.
\end{equation}
We compute the expected relative variance of the recovered \CII luminosity density, ${\rm var}(\rho_{\rm CII})/\rho_{\rm CII}$, as a function of $t_{\rm obs}$ for four different surveys with different telescope diameter $D$, survey area $\Omega_{\rm survey}$, or bandwidth. 
These cases allow us to compare the survey performances in terms of small and large diameters (S1 vs. S3), narrow and wide field (S1 vs. S2),  low and high angular resolution (S1 vs. S3), and number of bolometers and higher noise level (S4).
We use the L18 $L_{\rm CII}-{\rm SFR}$ relation to make the input signal. The details of each survey are listed in  Tab. \ref{tab:strategy}. 
The results (in units percentage) are plotted in Fig. \ref{fig:strategies}, where the relative $\rho_{\rm CII}$ variances obtained from the shot-noise only (power spectrum at $k_{\rm max}$), and from the full power spectrum are shown separately by thin/thick lines, respectively.  
 
Even if the shot-noise measurements could reach very high $S/N$, the derived \CII luminosity density still has large uncertainties, see the thin lines in Fig. \ref{fig:strategies}. 
This is not due to the instrumental noise, but rather to the intrinsic uncertainties introduced by $L_{\rm CII}-{\rm SFR}$ relation parameter degeneracies. In such a case, increasing the observational time yields only marginal improvements. 
If instead the full power spectrum is available, ${\rm var}(\rho_{\rm CII})/\rho_{\rm CII}$ is reduced significantly, since the clustering term helps to break the parameter degeneracies (see the thick lines in Fig. \ref{fig:strategies}).

The S1 survey can obtain the \CII luminosity density with relative uncertainties $\sim40\%$. S3, which has the same survey volume as S1 but uses a larger telescope, does not increase the precision on $\rho_{\rm CII}$. On the other hand,  S2, using the same telescope as S1 but a larger volume in the same observation time, yields a larger 
 ${\rm var}(\rho_{\rm CII})/\rho_{\rm CII}$ than S1. In brief, a S1-like survey is recommended based on our study.  Finally, a mini-survey (S4) which could be readily performed with current instruments would constrain var($\rho_{\rm CII})/\rho_{\rm CII}$ to $\sim800\%$ if  $\sim1000$ hr of observation time and $\sim3.8$ GHz bandwidth are dedicated to a relatively small area of $\rm (0.2~deg)^2$.

Instead of just observing a single field, the survey might include several independent fields fairly far from each other on the sky. 
This strategy could reduce the cosmic variance. However, using the \citet{Driver2010} cosmic variance calculator\footnote{\url{https://rdrr.io/cran/celestial/man/cosvar.html}}, we find that for a field with area 4 deg$^2$ and depth from $z=6-7$, the cosmic variance is $\lesssim10\%$. This does not represent a major problem for a high-$z$ \CII IM survey. Another advantage of a multi-field survey is that at different sky positions the Galactic foreground can be very different. Thus, by comparing the \CII signal in different fields, one can check the influence of residual Galactic foregrounds.

\begin{table*}
\caption{Comparison between different survey strategies}
\begin{center}
\begin{tabular}{|c|c|c|c|c|c|c|c|c|c|c|c|c|c|c|c|c|c|c|}
\hline\hline
Name &&S1$^*$&S2&S3&S4 \\
\hline
D & [m] & 6 & 6& 25& 6\\
$N_{\rm det}$& \#& 4000&4000 & 4000& 400\\
$\delta \nu_0$ & [GHz]& 1.0 & 1.0& 1.0 &0.1\\
$T_{\rm sys}$ & [K] & 50&50&50 &150 \\
NEFD & Jy sr$^{-1}$ s$^{1/2}$ & $2.4\times10^6$ & $2.4\times10^6$& $2.4\times10^6$ & $2.3\times10^7$ \\
Band & [GHz] & $237.6-271.5$ & $237.6-271.5$ & $237.6-271.5$ & $267.7-271.5$ \\
Bandwidth & [GHz] & 33.9 & 33.9 &33.9 & 3.8\\
Redshift & & $6.0-7.0$ & $6.0-7.0$&$6.0-7.0$&$6.0-6.1$\\
Survey area & [deg$^2]$ & $4.0$ &$100.0$& $4.0$ & $0.04$ \\
Beam$^\dag$ & [arcsec] &46.3 &46.3 & 11.1&46.3 \\
Survey volume & [Mpc$^3$] & $3.2\times10^7$ & $8.1\times10^8$ &$3.2\times10^7$& $3.7\times10^4$\\
$V_{\rm vox}$ & [Mpc$^3]$ & 39.6 & 39.6&2.3 & 4.0 \\
$t_{\rm obs}$ & [hour] & $10-2000$ & $10-2000$ & $10-2000$&$10-2000$\\
Instrumental noise & [Jy sr$^{-1}$] & $1.8\times10^4( \frac{t_{\rm obs}}{\rm 1000hr} )^{-1/2}  $ &$9.1\times10^4 ( \frac{t_{\rm obs}}{\rm 1000hr} )^{-1/2}  $ & $7.6\times10^4 ( \frac{t_{\rm obs}}{\rm 1000hr} )^{-1/2}  $  & $5.8\times10^4 ( \frac{t_{\rm obs}}{\rm 1000hr} )^{-1/2}  $\\
\---& [mJy beam$^{-1}$] & $0.92 ( \frac{t_{\rm obs}}{\rm 1000hr} )^{-1/2}$    & $4.6( \frac{t_{\rm obs}}{\rm 1000hr} )^{-1/2}$&$0.2 ( \frac{t_{\rm obs}}{\rm 1000hr} )^{-1/2}$& $2.9( \frac{t_{\rm obs}}{\rm 1000hr} )^{-1/2}$\\
$k_{\rm max}$ & $[{\rm Mpc^{-1}}]$& $1.7$ & $1.7$ & $6.9$ & $3.3$ \\
\hline
\end{tabular}
\end{center}
$^*$ S1 (for $t_{\rm obs}=1000$ hour) is our fiducial survey. \\ 
$^\dag$ At 271.5 GHz. \\
\label{tab:strategy}
\end{table*}

\section{Contamination}\label{sec:foreground}

So far we have ignored the continuum foreground and intervening lines, assuming they could be perfectly removed and only \CII signal leaves. It is beyond the scope of this work to explore in detail foreground removal algorithms. We limit ourselves to present the level of continuum foregrounds from different sources and compare them with the \CII signal, from which one can preliminary deduce the requirements on the residual foreground levels.  We also simply test the feasibility of two intervening line removal methods.

\subsection{Continuum foregrounds}

The continuum foregrounds include the CIB, Galactic dust emission and CMB. The specific CIB intensity is
\begin{equation}
I_{\rm CIB}(\nu_0)=\frac{1}{4\pi}\int \frac{cdz}{H(z)(1+z)} l_{\nu_0(1+z)}(z) {\rm SFRD}(z),
\end{equation}
where $l_{\nu_0(1+z)}$ is the redshift-evolved mean galaxy SED template \citep{Bethermin2012,Magdis2012} and ${\rm SFRD}(z)$ is the cosmic star formation rate density constrained by observed IR and UV luminosity densities. We use \citep{Puchwein2019} \begin{equation}
{\rm SFRD}(z)=0.01\frac{(1+z)^{2.7}}{1+[(1+z)/3.0]^{5.35}}~M_\odot {\rm yr^{-1} Mpc^{-3}}.
\end{equation}

The Galactic dust emission follows a modified black-body spectrum \citep{Penin2012}
\begin{equation}
I_{\rm cirrus}(\nu_0)=I_0\left(\frac{\nu_0}{3000{\rm GHz}}\right)^{2}B_{\nu_0}(15.9{\rm K}),
\end{equation}
where $B_{\nu_0}$ is the black-body spectrum.  The normalization factor $I_{\nu_0}$ is determined by  $I_{\rm cirrus}(3000{\rm GHz})=1.25\times10^6$ Jy/sr, as measured in a low-dust field ELAIS N1 \citep{Penin2012}. The Lockman Hole filed has even smaller Galactic dust emission, say $0.51\times10^6$ Jy/sr at 3000 GHz \citep{Lagache2007}. Such low-dust ``holes" are recommended for IM experiments. 

\begin{figure}
\subfigure{\includegraphics[width=90mm]{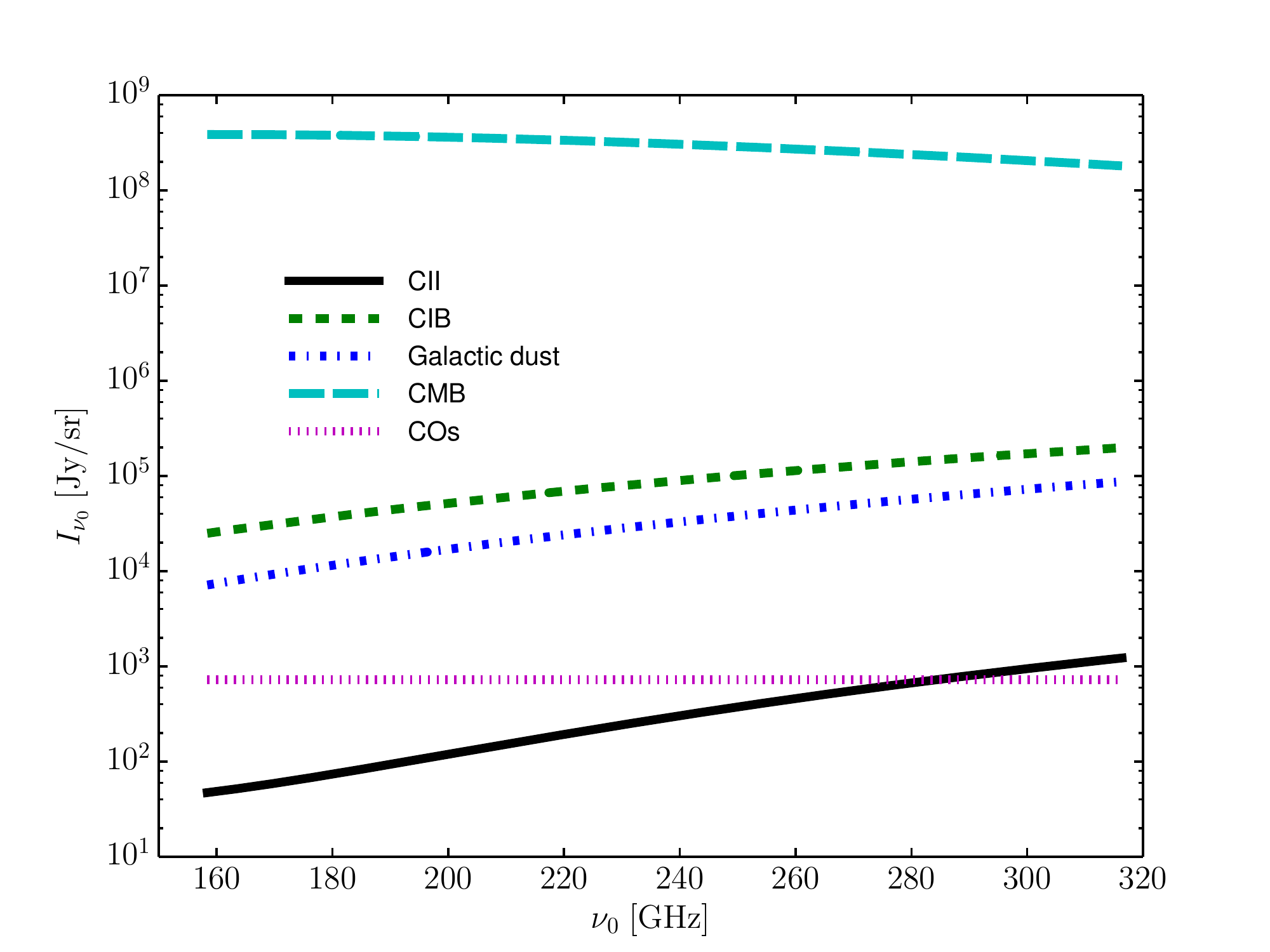}}
\caption{
The absolute specific intensity of the \CII emission, CIB, Galactic dust emission, and CMB in the frequency range corresponding to the \CII signal from $z=10-5$.
}
\label{fig:foreground_flux}
\end{figure}

In Fig. \ref{fig:foreground_flux} we plot the absolute specific intensity of the CIB, the Galactic dust emission and the CMB respectively, together with the \CII signal in deL14d model. Obviously, all foregrounds are much larger than the \CII signal, and the CMB is the strongest one.

We then go to the foreground fluctuations. Since all foregrounds are continuum in frequency space, in principle they only have angular fluctuations. We therefore compare their angular power spectra with the angular power spectrum of a 2D \CII IM slice with central frequency $\nu_0$ and thickness $\delta \nu_0$. The projected angular power spectrum of this slice \citep{Neben2017} is
\begin{equation}
P^{\rm 2D}_{\rm CII}(q)=\frac{1}{r^2\delta r}P_{\rm CII}\left(k=\frac{q}{r},z\right),
\end{equation}
where $q=2\pi/\theta$ is the angular wavenumber corresponding to an angle $\theta$; $\delta r=c/H(z)\delta \nu_0(1+z)/\nu_0$ is the comoving thickness  of the slice. 

Regarding the CIB, we use the angular power spectrum measured by  {\tt Planck} at 353 GHz \citep{Lenz2019} and rescale it to $\nu_0$ by 
using the ratio $[I_{\rm CIB}(\nu_0)/I_{\rm CIB}(353{\rm GHz})]^2$.

\citet{Penin2012} measured the angular power spectrum of the Galactic dust emission in the ELAIS N1 field,
\begin{equation}
P^{\rm 2D}_{\rm cirrus}(q)=4.93\times10^6\left( \frac{q}{0.01\times2\pi~{\rm arcmin^{-1}}}  \right)^{-2.53}~{\rm (Jy/sr)^2 sr}
\end{equation}
at 3000 GHz; we rescale it to $\nu_0$ by the ratio $[I_{\rm cirrus}(\nu_0)/I_{\rm cirrus}(3000{\rm GHz})]^2$.

For the CMB power spectrum, we directly use the best-fit of the {\tt Planck}2015 TT measurements \citep{Planck2015cos}, and convert it into the specific intensity fluctuations at $\nu_0$.

\begin{figure}
\subfigure{\includegraphics[width=90mm]{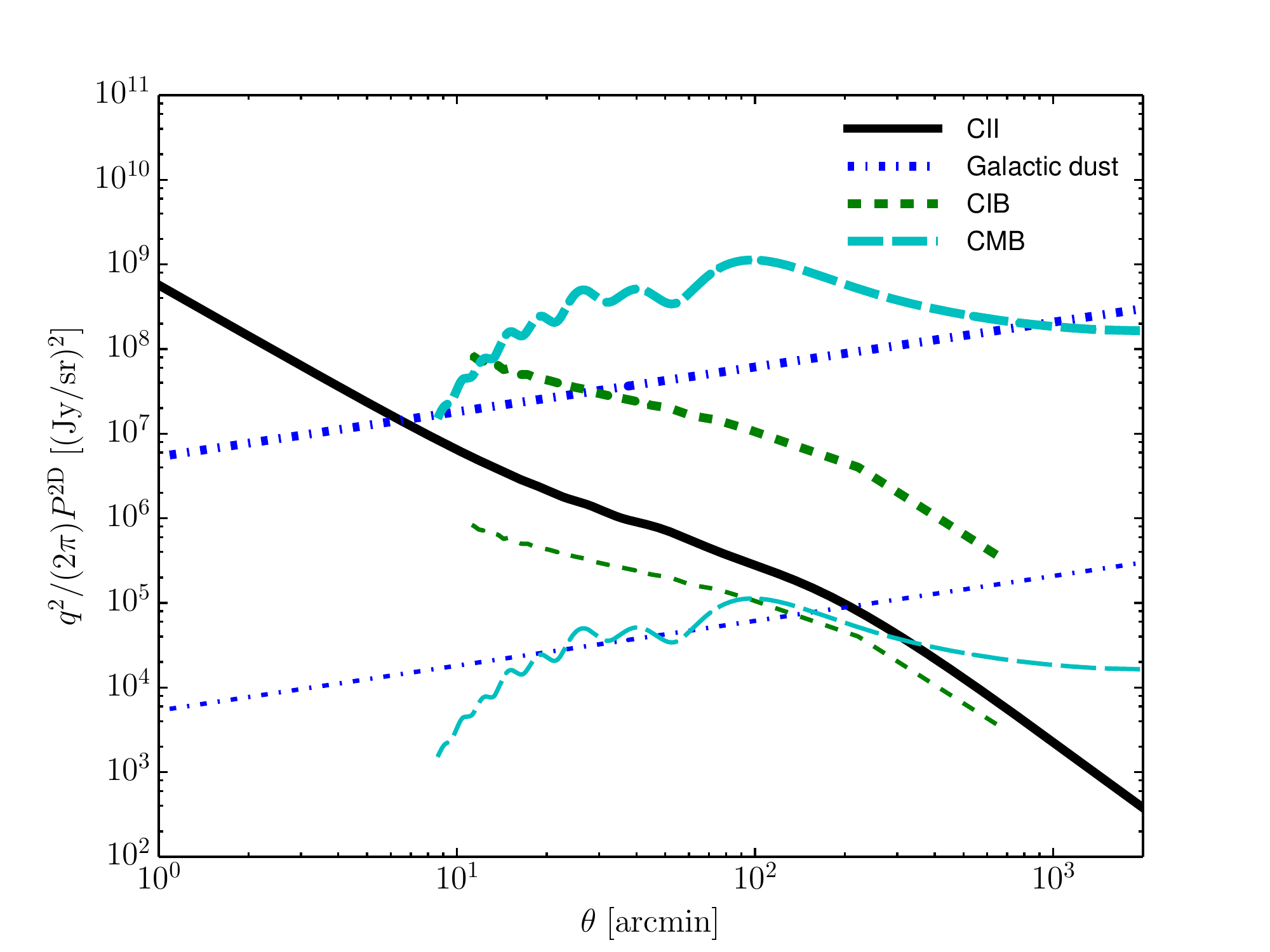}}
\caption{
 The angular power spectrum of the \CII emission, the CIB, the Galactic dust emission and the CMB at $\nu_0=271.4$ GHz (thick lines). Thin lines refer to 0.1\% of the CIB and Galactic dust emission power spectra, and 0.01\% of the CMB power spectrum. 
 }
\label{fig:foreground_fluc}
\end{figure}

In Fig. \ref{fig:foreground_fluc} we plot the angular power spectra of the \CII signal from $z\sim6$, the CIB, the Galactic dust emission and CMB at the same frequency, respectively. Since foregrounds are smooth in frequency, in principle most of them could be extracted from the map. For example, for the CMB many tools are available for components separation based on multi-color maps, e.g., \citet{GNILC,ABS_Zhang}. In Fig. \ref{fig:foreground_fluc} we also plot 1\% of the CIB power, 0.1\% of the Galactic dust emission power, and 0.01\%  of the CMB power (the residual flux levels are $\sqrt{1\%}=10\%$, $\sqrt{0.1\%}\approx3\%$ and $\sqrt{0.01\%}\approx1\%$ respectively). They are roughly comparable with the \CII signal, thus these values set the limiting requirements on the residual foreground.

\subsection{Interloping lines}

\begin{figure}
\subfigure{\includegraphics[width=90mm]{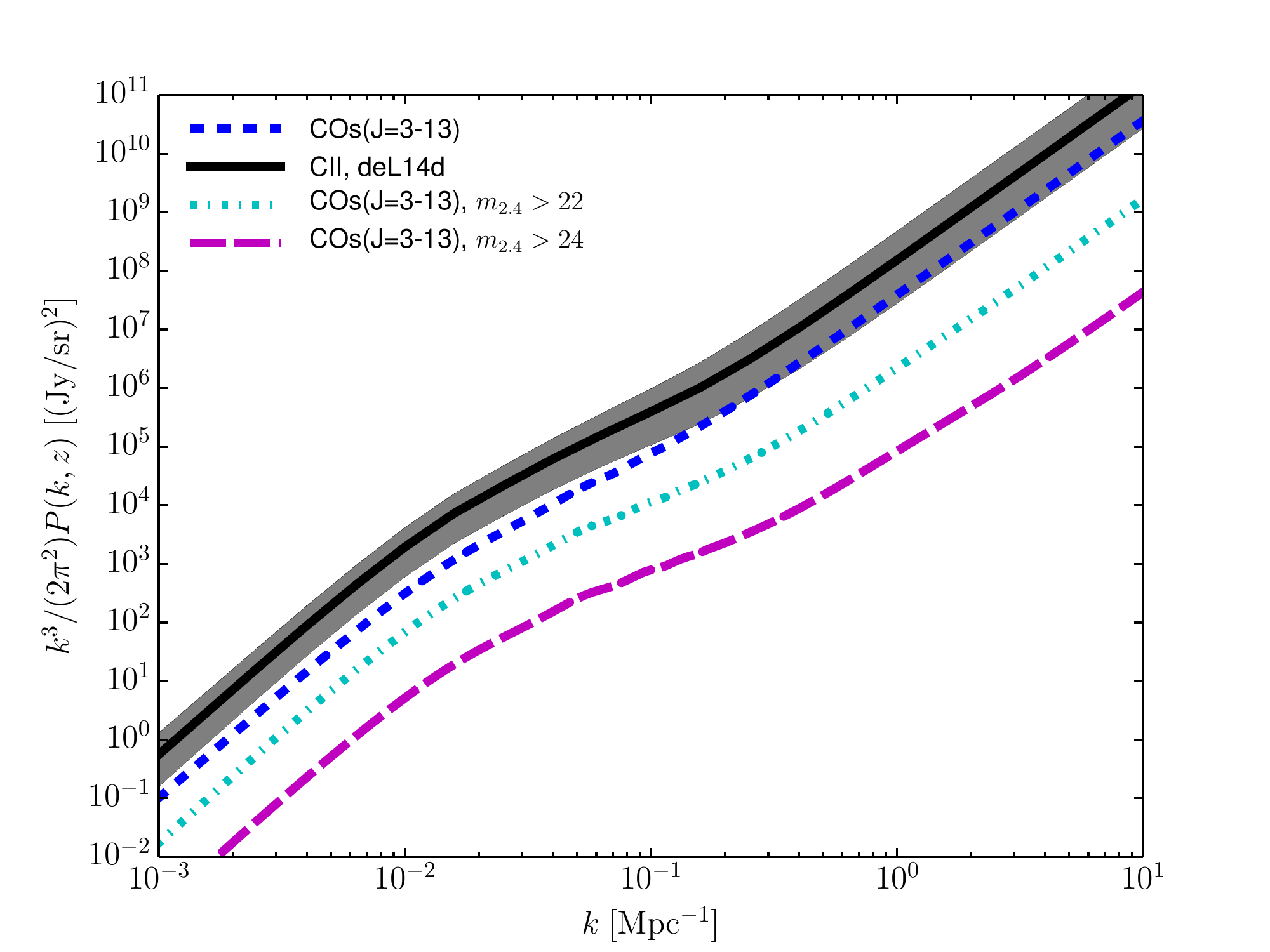}}
\caption{
The angular power spectrum of the \CII emission from $z\sim6$ and the sum of CO lines redshifted into the same frequency.
 }
\label{fig:interloping}
\end{figure}

 For \CII signal detected at $\nu_0$,  interloping lines are basically CO lines with rotational transitions with quantum number $J$ from $z=\nu_{\rm CO,J}/\nu_0-1$ as long as $\nu_{\rm CO,J} > \nu_0$. For example for \CII signal from $z\sim6$, $J\ge3$. 
Analogous to \CII, using the relation between CO luminosity and IR luminosity for $J=1-13$ derived from observations of {\tt Herschel} and ground-based telescopes \citep{Greve2014}, and the IR LF \citep{Gruppioni2013},  we derive the absolute specific intensity and fluctuation amplitude of CO lines. The absolute specific intensity of COs for all observable $J$ values as a function of the observed frequency is plotted in Fig. \ref{fig:foreground_flux}. The power spectrum of lines that could contaminate $z\sim6$ \CII signal, i.e., $J=3-13$, is shown in Fig. \ref{fig:interloping}. In this case, CO interlopers are not a severe problem since their power is well below the \CII signal. However, as both the \CII and CO models are highly uncertain, the interloping lines removal is necessary. Some algorithms have been discussed in, e.g. \citet{Gong2014,Silva2015,Yue2015,Breysse2015,Lidz2016,Cheng2016,Sun2018}.
Basically they either (a) remove bright foreground voxels/galaxies, or (b) take advantage of the asymmetry of the low-$z$ CO power spectra parallel and perpendicular to the line-of-sight.

Using the mean SED of submillimeter galaxies in \citet{Michalowski2010}, adding a 0.2 dex scatter on the CO luminosity - IR luminosity relation, we get the relation between the CO luminosity and apparent AB magnitude at 2.4 $\mu$m, $m_{2.4}$, to finally get the power spectrum of CO lines contributed by galaxies $>m_{2.4}$. In Fig. \ref{fig:interloping} we plot the power spectrum of CO lines with J=$3-13$, in which all galaxies brighter than AB=22 or 24 are removed. This algorithm effectively suppresses the CO interloping lines power.

Another method to separate the CO interloping lines from the \CII signal relies on the fact that interloping lines come predominantly from low-$z$ galaxies. As such, the power spectrum is asymmetric in modes parallel ($k_\parallel$) and perpendicular ($k_\perp$) to the line-of-sight. This method is described in detail in the literature \citet{Gong2014,Lidz2016}, and has been widely applied.

The cylindrical power spectrum of the emission line signal is written as \citet{Lidz2016}
\begin{equation}
P^{\rm cylin}_{\rm line}(k_\parallel,k_\perp,z_{\rm line})=I^2_{\rm line}\mean{b}^2_{\rm line}(1+\beta_{\rm line}\mu^2)^2P(k,z_{\rm line})+P_{\rm SN},
\label{eq:P_cyline_line}
\end{equation}
where $k=\sqrt{k_\parallel^2+k_\perp^2}$, $\mu=k_\parallel/k$ and $\beta=\frac{1}{\mean{b}_{\rm line}}\frac{d{\rm ln}D }{d{\rm ln}a}$ is the logarithmic derivative of the linear growth factor divided by mean emission line bias $\mean{b}_{\rm line}$. 

\begin{figure*}
\subfigure{\includegraphics[width=0.33\textwidth]{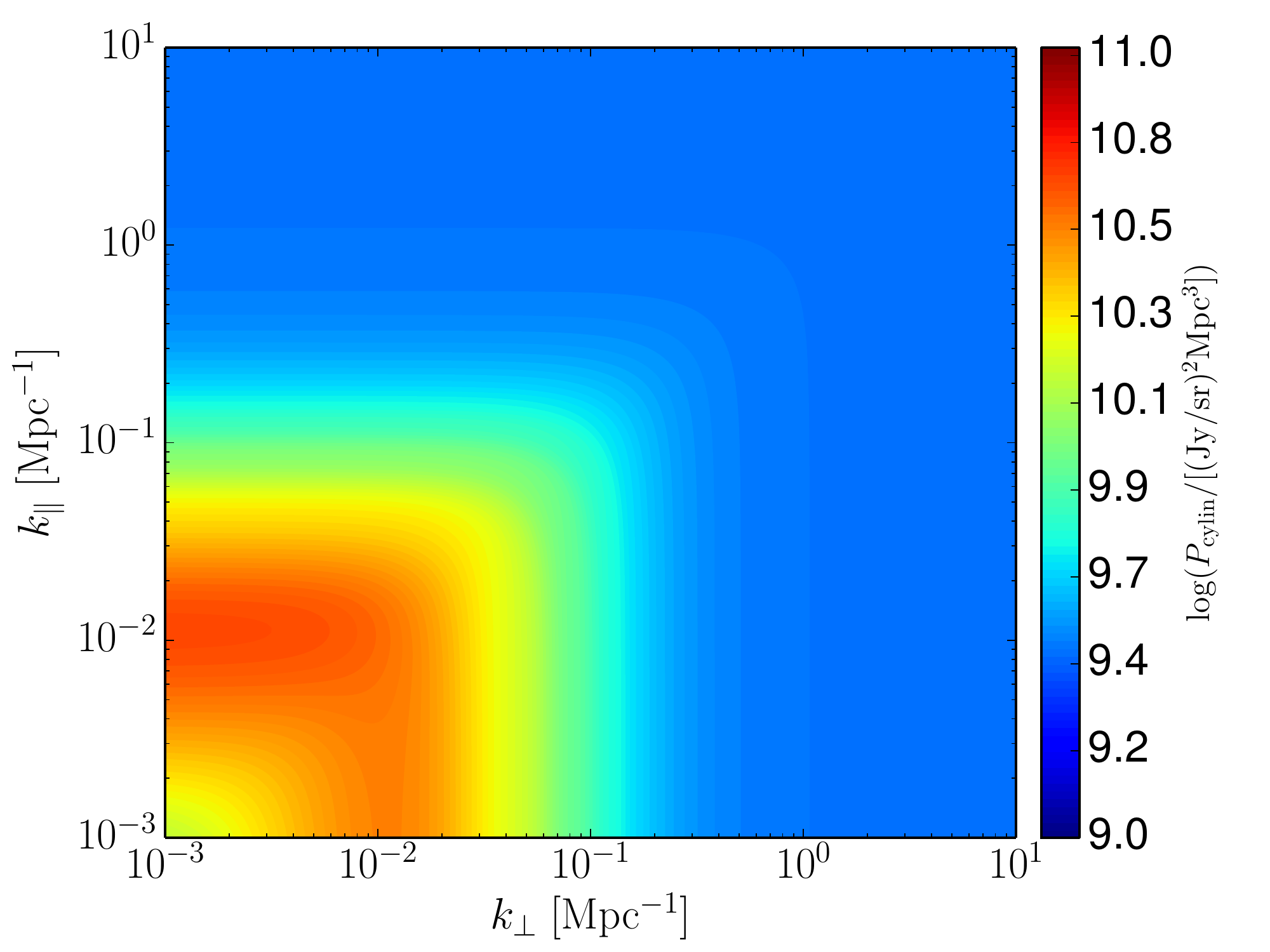}}
\subfigure{\includegraphics[width=0.33\textwidth]{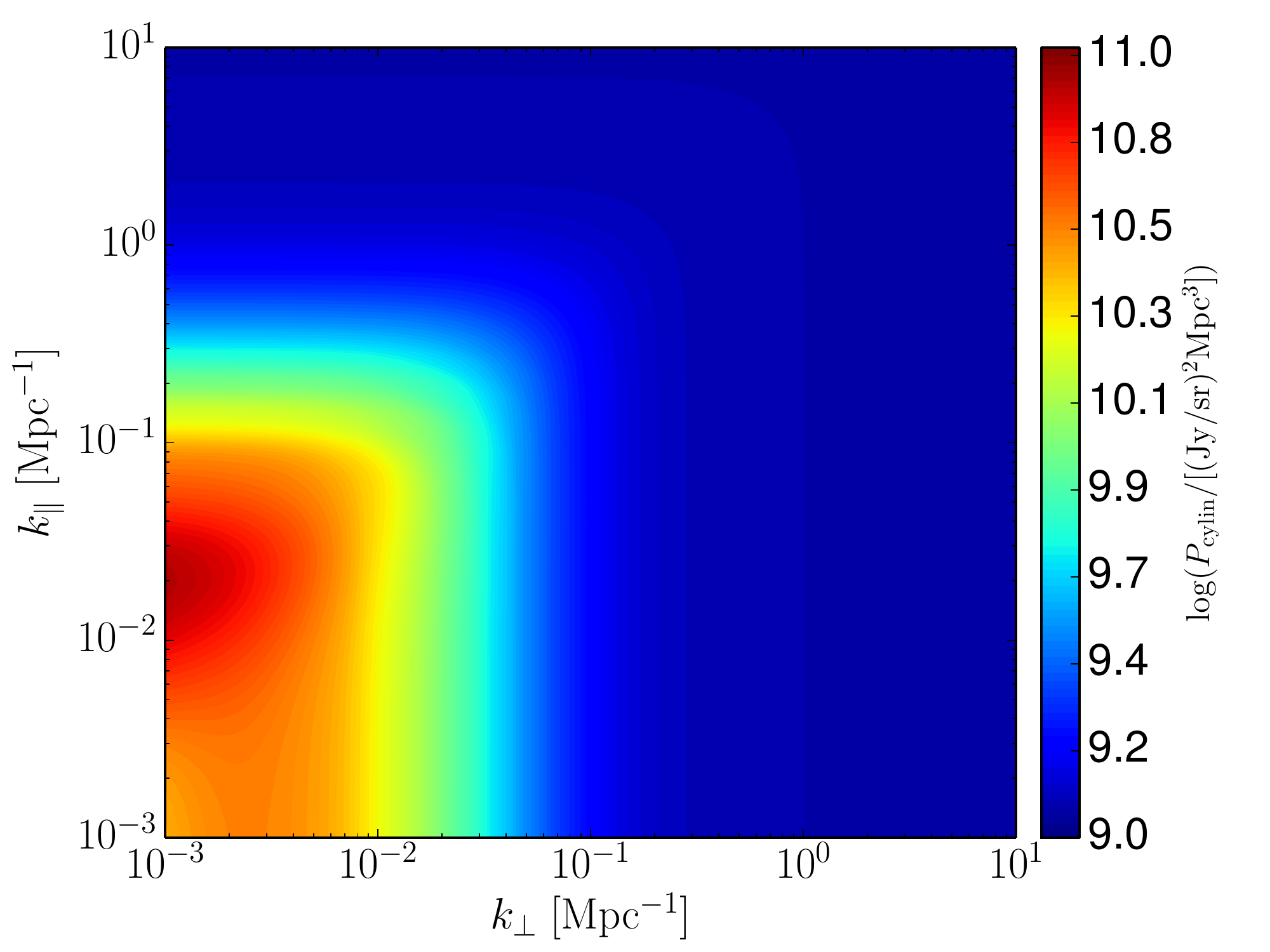}}
\subfigure{\includegraphics[width=0.33\textwidth]{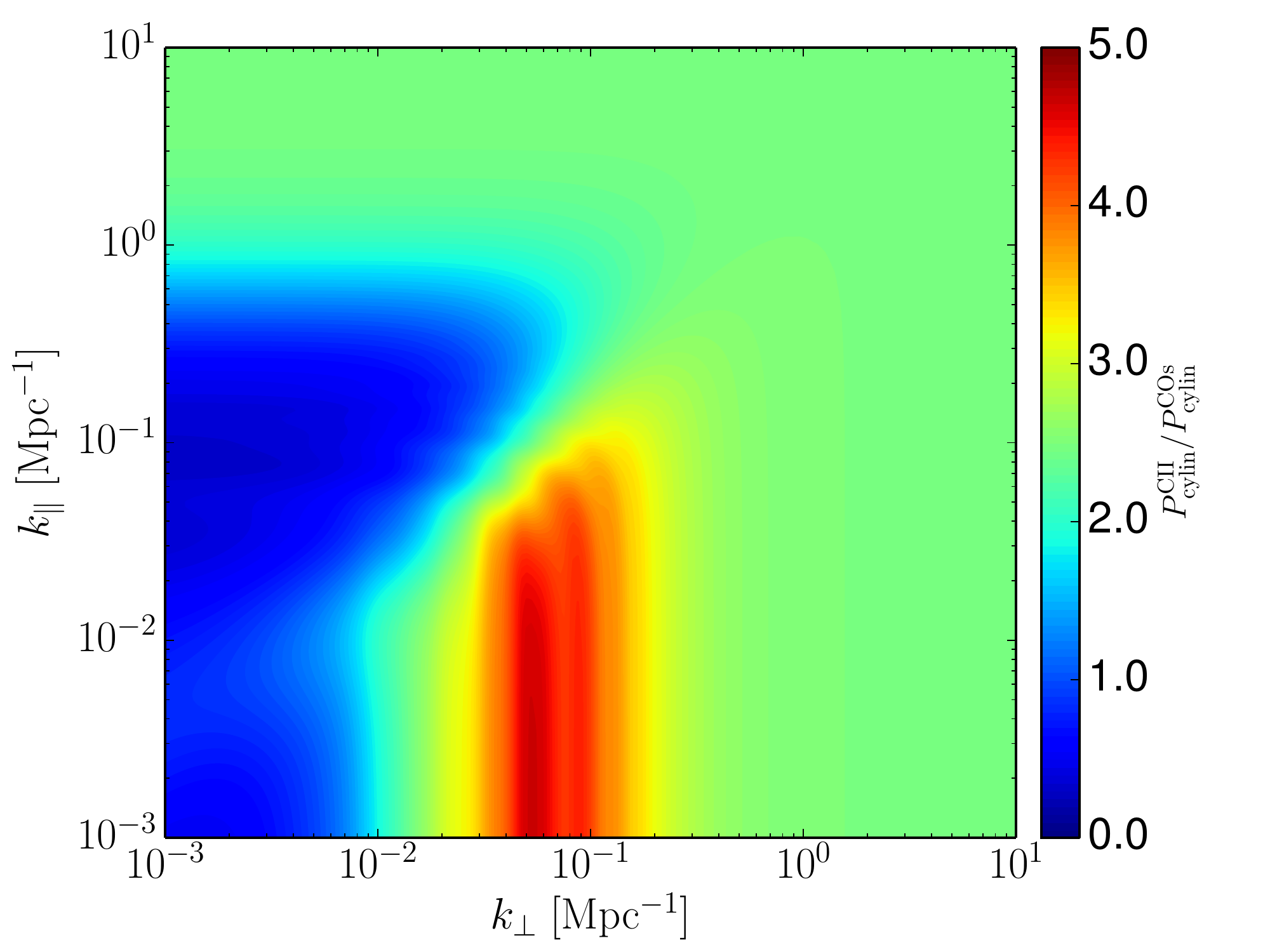}}
\caption{
The cylindrical power spectrum of \CII signal from $z\sim6$ (left), the apparent cylindrical power spectrum of CO interloping lines for $J=3-13$ which could contaminate the \CII signal (middle), and the \CII to CO line ratio (right).
 }
\label{fig:interloping_cylin}
\end{figure*}

However, if an interloping line from $z_{\rm loper}$ is incorrectly treated as signal from $z_{\rm line}$, then the measured cylindrical power is
\begin{equation}
\tilde{P}_{\rm loper}^{\rm cylin}(k_\parallel,k_\perp,z_{\rm loper},z_{\rm line})=\frac{1}{\alpha_\parallel \alpha^2_\perp}P^{\rm cylin}_{\rm loper}\left(\frac{k_\parallel}{\alpha_\parallel},\frac{k_\perp}{\alpha_\perp},z_{\rm loper}\right),
\end{equation}
where 
\begin{equation}
\alpha_\parallel=\frac{H(z_{\rm line})(1+z_{\rm loper})}{H(z_{\rm loper})(1+z_{\rm line})},
\end{equation}
and 
\begin{equation}
\alpha_\perp=\frac{r(z_{\rm loper})}{r(z_{\rm line})}.
\end{equation}
$P_{\rm loper}^{\rm cylin}$ is analogous to Eq. (\ref{eq:P_cyline_line}), except that all line variables are replaced with interloper variables. The measured total power spectrum   $P_{\rm tot}^{\rm cylin}=P_{\rm line}^{\rm cylin}+\tilde{P}_{\rm loper}^{\rm cylin}$.
For \CII signal and CO interloping lines corresponding to spherical power spectra shown in Fig. \ref{fig:interloping}, we plot the cylindrical power spectra, and the \CII to CO lines ratio in Fig. \ref{fig:interloping_cylin}.
The asymmetry is obviously seen in the ratio panel, since for low-$z$ interloping lines the scale is compressed along the line-of-sight while stretched perpendicular to line-of-sight. Kaiser effect also contributes part to this asymmetry.

We then find constraints on model parameters $I_{\rm CII}$, $I_{\rm CO}$, $P_{\rm SN}^{\rm CII}$ and $P_{\rm SN}^{\rm CO}$ by fitting both the \CII signal and CO interloping lines simultaneously. For simplicity we only consider a single J=4 CO line, i.e. the strongest interloping line. We also fix the bias, and assume our S1 survey. In this case the uncertainty on the cylindrical power spectrum is 
\begin{equation}
\Delta P_{\rm tot}^{\rm cylin}=\frac{1}{\sqrt{N_m(k_\parallel,k_\perp)}}( P_{\rm tot}^{\rm cylin}+P_N  ),
\end{equation}
where
\begin{equation}
 N_m(k_\parallel,k_\perp)=(\pi k^2_\perp d{\rm ln}k)(  k_\parallel d{\rm ln}k) \frac{V_{\rm survey}}{(2\pi)^3}
\end{equation}
is the number of $k$ modes in the relevant bins. 

\begin{figure}
\subfigure{\includegraphics[width=90mm]{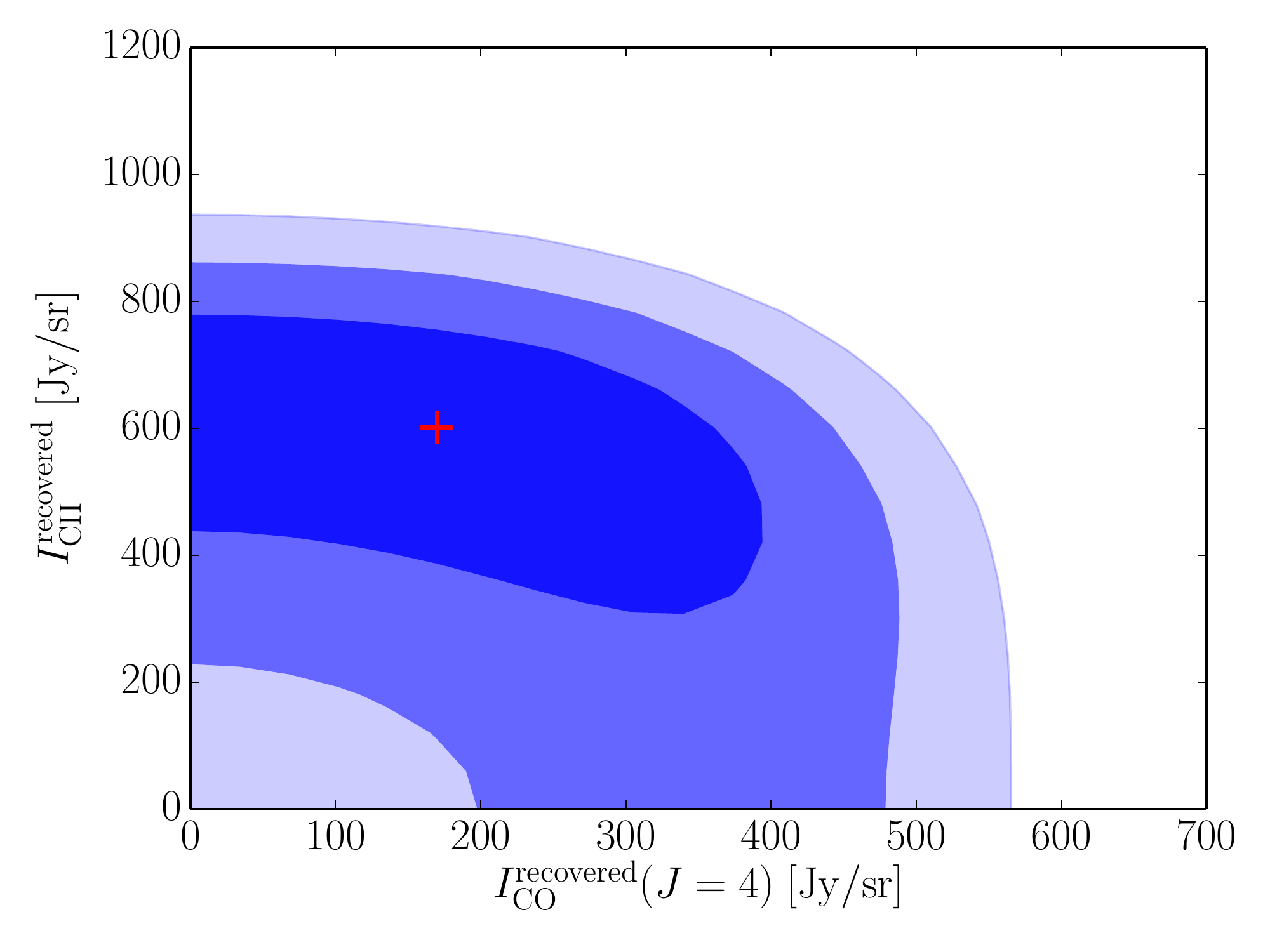}}
\caption{The constraints on the \CII signal and CO(J=4) interloping line for a survey of S1. From inside-out, the filled regions correspond to 1, 2 and 3 $\sigma$ constraints respectively. The input values is marked by a cross.
 }
\label{fig:chi2_ICII_ICO}
\end{figure}

\begin{figure}
\subfigure{\includegraphics[width=90mm]{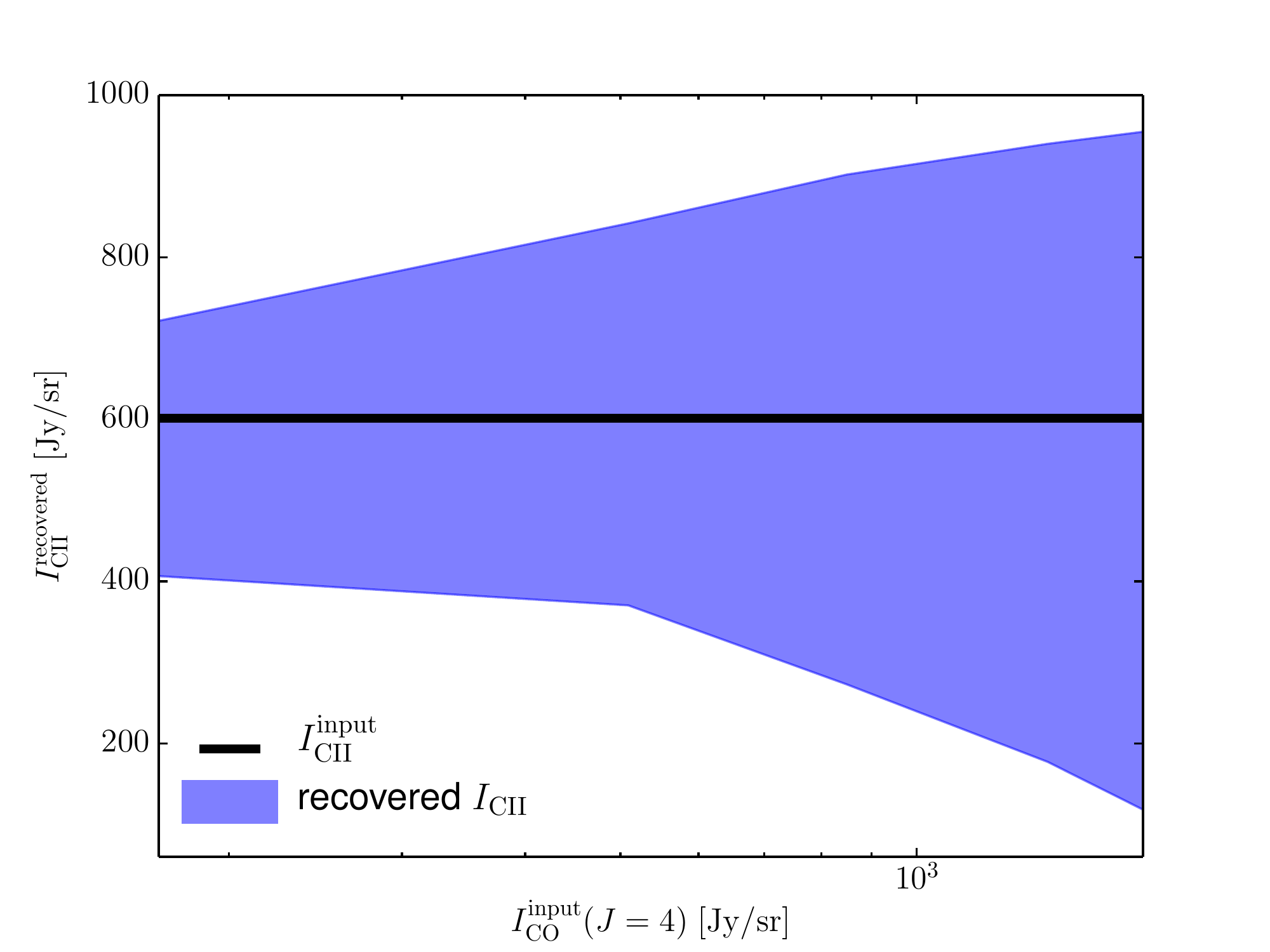}}
\caption{The recovered $I_{\rm CII}$ as a function of input CO interloping line specific intensity.
 }
\label{fig:ICII_constraint}
\end{figure}

The constraint on $I_{\rm CII}$ and $I_{\rm CO}$ (we have marginalized on $P_{\rm SN}^{\rm CII}$ and $P_{\rm SN}^{\rm CO}$) are shown in Fig. \ref{fig:chi2_ICII_ICO}, where the input value is marked by a cross. Further marginalizing the $I_{\rm CO}$, we have $I_{\rm CII}=580_{-180}^{+130}$ Jy/sr, while the input \CII specific intensity is 601 Jy/sr. To further test the performance of the scheme for stronger CO interloping lines, we also artificially boost the CO contribution, and show the final recovered \CII intensity as a function of input CO line intensity in Fig. \ref{fig:ICII_constraint}.  As the CO level increases, the recovered \CII signal shows larger uncertainties. 

\section{Conclusions}
\label{SandD}

We have investigated the possibility to study high redshift galaxies via \CII IM experiments. Our results clarify the conditions under which an IM experiment can be used to: (a) detect galaxies fainter than currently observed by pointed observations;  (b) detect rare, massive sources that cannot be traced in small fields as the { ALMA}/{\it HST} ones but which could be studied in the future with, e.g., {\it Euclid}/{\it WFIRST}. 

To address these issues, we have envisaged a fiducial IM survey performed with a dish telescope of diameter $D=6$ m,  $T_{\rm sys}=50$ K, $N_{\rm det}=4000$ bolometers and frequency resolution $\Delta \nu_0=1$ GHz.  These specifications are similar to (and inspired by) the forthcoming CCAT-p telescope. The assumed survey angular area is $(2~{\rm deg})^2$ at $z=6$ with depth $\Delta z=1$, using observation time $t_{\rm obs}=1000$ hr. 

We derived the $z\sim6$ \CII LF from the observed UV LF, assuming a form ${\rm log}L_{\rm CII}={\rm log}A+\gamma{\rm SFR}\pm \sigma_L$. We have allowed log$A$ and $\gamma$ to vary in a broad range, including estimates from theoretical and observational studies from literatures, and calculated the IM power spectrum and its signal-to-noise ratio. 
We mainly concentrated on the detectability of the \CII power spectrum by IM experiments, and the possibility to recover the \CII LF either from the shot-noise alone or from the full spectrum. 

The main results are the following: 
\begin{itemize}
\item 
In our fiducial survey (inspired by CCAT-p/1000hr) at $z\sim6$ the shot-noise (clustering) signal is detectable for 2 (1) of the 5 considered $L_{\rm CII}$ – SFR relations from literatures.

\item
With the only exception of the O17 relation, the shot-noise is dominated by galaxies with $L_{\rm CII}\gtrsim 10^{8-9}~L_\odot$, already well at reach of {ALMA}. However, the IM experiment can  provide crucial information on the bright-end of the LF.

\item If the $L_{\rm CII}-{\rm SFR}$ relation varies in log$A-\gamma$ wider than the 5 considered $L_{\rm CII}-{\rm SFR}$ relations, but yet still consistent with upper limits from \CII blind searches, the signal produced by galaxies even as faint as $L_{\rm CII}\sim 10^7~L_\odot$ may be detectable. Hence the detection depth of an IM experiment crucially depends on the $L_{\rm CII}-{\rm SFR}$ relation in the EoR that is yet poorly constrained. 

\item The detection of the \CII power spectrum signal allows to reconstruct the \CII LF. 
Although the shot-noise has higher $S/N$ than the clustering, it can not put separate constraints on the log$A$ and $\gamma$ parameters that determine the LF. The full power spectrum is crucial to break the degeneracy between log$A$ and $\gamma$ and precisely recover the LF.

\item We considered four possible survey strategies,  whose parameters are reported in Tab. \ref{tab:strategy}, and discussed their pros and cons in terms of deriving the \CII luminosity density.  

\end{itemize} 
\label{Con}
 
We conclude with a caveat. In this paper we have assumed  foreground- and contamination-cleaned \CII~signal. 
A method for foreground and contamination subtraction has been discussed in \citet{Yue2015}. Here we devoted some discussions to the contamination by continuum foregrounds such as the  CIB, Galactic dust emission, and CMB, along with the requirements on the  residual contamination level to reliably extract the \CII signal. We also analyzed the contamination by CO interloping lines, and tested two interloper removal methods. We conclude that the two proposed methods successfully recover the \CII signal, at least under our testing conditions.
One could also go around the foreground problem with cross-correlations with other IM data, such as CO or HI. However, since the foreground is much larger than the \CII signal, even a tiny residual foreground might represent a problem. For integration times long enough that residual foreground -- rather than instrumental noise  -- dominates the error budget, a further decrease of $P_N$ would bring no advantage. Instead, increasing the survey area would be useful because  the residual foreground and contamination angular power spectrum could be different from the \CII~signal on large scales.

\section*{Acknowledgments} 
We thank the anonymous referee for the very useful comments and suggestions. 
BY acknowledges the support of  the CAS Pioneer Hundred Talents (Young Talents) program, the NSFC grant 11653003, the NSFC-CAS  joint fund for space scientific satellites No. U1738125, and the NSFC-ISF joint research program No. 11761141012.
AF acknowledges support from the ERC Advanced Grant INTERSTELLAR H2020/740120. This research was supported by the Munich Institute for Astro- and Particle Physics (MIAPP) of the DFG cluster of excellence ``Origin and Structure of the Universe".
BY \& AF thank NORDITA for its hospitality during the NORDITA 2019 program  ``Zoom-in and Out: from the Interstellar Medium to the Large Scale Structure of the Universe".
 
\bibliographystyle{mn2e}
\bibliography{paper}

\newpage 
\label{lastpage} 
\end{document}